%% file: radar20.tex
\documentclass[conference]{IEEEtran}

\usepackage[utf8]{inputenc} 
\usepackage[T1]{fontenc}    
\usepackage{url}            
\usepackage{booktabs}       
\usepackage{amsfonts}       
\usepackage{nicefrac}       
\usepackage{microtype}      

\usepackage{pgfplots}
\usetikzlibrary{positioning,shapes,arrows,patterns,fit,backgrounds,calc,svg.path,3d,decorations.text,shapes.arrows}
\usepackage{todonotes}
\usepackage{graphicx}
\usepackage{subfigure}
\usepackage{amsmath,amssymb}
\usepackage{array}
\usepackage{makecell}
\usepackage{caption}

\pgfdeclarelayer{background}
\pgfdeclarelayer{front}
\pgfsetlayers{background,main,front}

\makeatletter
\pgfkeys{%
	/tikz/on layer/.code={
		\pgfonlayer{#1}\begingroup
		\aftergroup\endpgfonlayer
		\aftergroup\endgroup
	},
	/tikz/node on layer/.code={
		\gdef\node@@on@layer{%
			\setbox\tikz@tempbox=\hbox\bgroup\pgfonlayer{#1}\unhbox\tikz@tempbox\endpgfonlayer\egroup}
		\aftergroup\node@on@layer
	},
	/tikz/end node on layer/.code={
		\endpgfonlayer\endgroup\endgroup
	}
}

\def\node@on@layer{\aftergroup\node@@on@layer}

\pgfplotsset{ every non boxed x axis/.append style={x axis line style=-},
	every non boxed y axis/.append style={y axis line style=-}}

\def\BibTeX{{\rm B\kern-.05em{\sc i\kern-.025em b}\kern-.08em
		T\kern-.1667em\lower.7ex\hbox{E}\kern-.125emX}}

\begin{document}
\title{Deep Interference Mitigation and Denoising of Real-World FMCW Radar Signals}

\author{
	\IEEEauthorblockN{Johanna Rock$^{1}$, Mate Toth$^{1,2}$, Paul Meissner$^2$,  Franz Pernkopf$^1$}
	\IEEEauthorblockA{$^1$Graz University of Technology, Austria
	$^2$Infineon Technologies Austria AG, Graz}
	Email: johanna.rock@tugraz.at
}


\maketitle

\begin{abstract}
Radar sensors are crucial for environment perception of driver assistance systems as well as autonomous cars. Key performance factors are a fine range resolution and the possibility to directly measure velocity. With a rising number of radar sensors and the so far unregulated automotive radar frequency band, mutual interference is inevitable and must be dealt with. Sensors must be capable of detecting, or even mitigating the harmful effects of interference, which include a decreased detection sensitivity.
In this paper, we evaluate a Convolutional Neural Network (CNN)-based approach for interference mitigation on real-world radar measurements. We combine real measurements with simulated interference in order to create input-output data suitable for training the model. We analyze the performance to model complexity relation on simulated and measurement data, based on an extensive parameter search. Further, a finite sample size performance comparison shows the effectiveness of the model trained on either simulated or real data as well as for transfer learning. A comparative performance analysis with the state of the art emphasizes the potential of CNN-based models for interference mitigation and denoising of real-world measurements, also considering resource constraints of the hardware.
\end{abstract}


%
\IEEEpeerreviewmaketitle

\section{Introduction}

Automotive radar sensors are key elements of current driver assistance systems and autonomous driving applications. In the automotive context, \emph{frequency modulated continuous wave (FMCW)/chirp sequence (CS)} radars are prevalent. They transmit sequences of linear chirp signals in a shared and non-regulated spectrum. Ever larger \emph{radio frequency (RF)} transmit bandwidths are required to fulfill the demand on fine range resolution. Because of these larger bandwidths and because of a rising number of deployed radar sensors, mutual interference is becoming increasingly likely.

Non-coherent interference, in which radar sensors with non-identical transmit signal parameters interfere, is the most common form of mutual interference~\cite{TOT18}. This leads to a reduced object detection sensitivity~\cite{BRO07}. Therefore, interference mitigation is a crucial part of current and future radar sensors used in a safety context.

Several conventional signal processing algorithms have been proposed in order to mitigate mutual interference. The most basic method is to zero out the interference-affected signal samples. More advanced methods use nonlinear filtering in slow-time~\cite{WAG18}, iterative reconstruction using Fourier transforms and thresholding~\cite{MAR12}, estimation and subtraction of the interference component~\cite{BEC17}, or beamforming~\cite{Bechter2016}. Some machine learning techniques were discussed in the context of interference detection and classification in~\cite{Zhang2018a}.

\emph{Convolutional Neural Networks (CNNs)} have been successfully used for image denoising, e.g. in ~\cite{zhang2017beyond}. CNN-based interference mitigation and denoising methods presented in~\cite{Rock1907:Complex} can be applied to \emph{range-Doppler (RD)} maps. A two-channel representation of the complex spectrogram data (i.e. real and imaginary data) is used as network input. Experimental results show a strong denoising and interference mitigation capability in comparison to state-of-the-art signal processing algorithms, though evaluated only on simulated data. The applicability of these models for robust interference mitigation on real-world data has not been investigated so far.

Denoising Autoencoders (DAEs) \cite{vishwakarma} and Generative Adversarial Networks (GANs) \cite{8541507, DBLP:journals/corr/abs-1811-04678} have been used for RF signal denoising, applied either in time domain or in frequency domain. These models achieve promising denoising performance, however, they typically require more learnable parameters and more complex model structures and thus are less suitable for deployment on resource-restricted hardware. Residual learning \cite{zhang2017beyond, 8394255, jiang2019deep} has successfully been used for denoising of unstructured noise.

In this paper, we analyze the suitability of CNN-based models from~\cite{Rock1907:Complex} to perform interference mitigation and denoising on real-world radar measurements. Therefore an extensive measurement campaign in a typical inner-city road traffic scenario has been carried out. The interference is simulated for both, simulated object scenarios and real-world measurements.

Due to the absence of labeled object positions in the target RD maps, we use the \emph{cell averaging constant false alarm rate (CA-CFAR)} algorithm \cite{scharf1991statistical} to identify the most likely object positions. These positions are the basis for our performance comparison using the \emph{signal-to-interference-plus-noise ratio (SINR)}~\cite{performance-comparison-interf-mitigation}.

Main contributions of this paper are:
\begin{itemize}
	\item We consider real-world radar measurements combined with simulated interference for CNN-based interference mitigation and denoising of RD maps.
	\item We compare performance and model complexity on simulated and measurement data to show that already small models yield good results.
	\item We analyze the effect of finite sample sizes on model performance and robustness.
	\item We present numerical results using application-related performance metrics in a comparison with the state of the art, i.e. Zeroing, IMAT and Ramp filtering.
	\item We show that an excellent level of noise reduction and hence an improvement of detection sensitivity can be achieved on real-world measurements.
\end{itemize}

\section{Signal model}
\label{sec:sigmod}

\begin{figure}[tb]
	\centering
	\input{spchain_classical.tex}
	\caption{Block diagram of a basic FMCW/CS radar processing chain. Dashed boxes indicate the locations of optional interference mitigation steps, including CNN-based approaches (red) and classical methods (blue).}
	\label{fig:spchain_classical}
	\vspace{-5mm}
\end{figure}
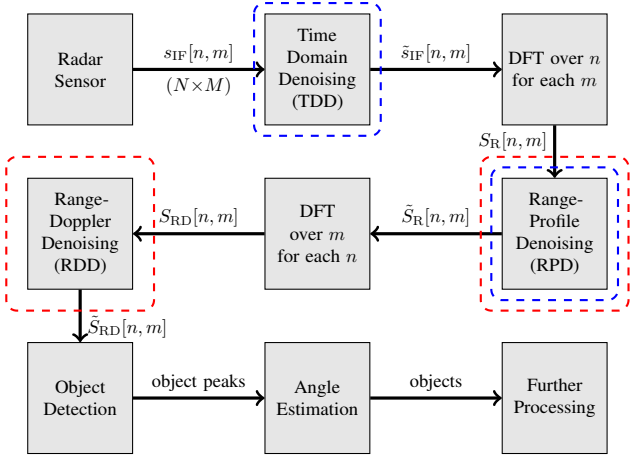

The \emph{range-Doppler (RD) processing} chain of a common FMCW/CS radar is depicted in Fig.~\ref{fig:spchain_classical}. A measurement is performed by transmitting a succession of $M$ linearly modulated RF chirp sequences. By mixing each chirp, also termed \emph{ramp}, with the received object reflections, the \emph{intermediate frequency (IF)} signal is obtained. It consists of sinusoidal components corresponding to objects. The objects' distances and velocities are contained in the sinusoidals' frequencies and their linear phase change over successive ramps~\cite{STO92,WIN07}, respectively.

We obtain $N$ samples per ramp; from a data processing point of view the IF signal can be interpreted as a $N \times M$ data matrix $s_{\mathrm{IF}}[n,m]$. Note that the indices $n$ and $m$ are commonly referred to as \emph{fast-} and \emph{slow-time}, respectively. Subsequently in the processing chain, discrete Fourier transforms (DFTs) are computed over both dimensions, yielding a two-dimensional spectrum on which peaks can ideally be found at positions corresponding to the objects' distances and velocities. After peak detection, further processing can include angular estimation, tracking, and classification.

Besides object reflections, real radar measurements may also include receiver noise and interference. In order to model these disturbances, we define the IF signal as

\begin{equation}
s_{\mathrm{IF}}[n,m]=\sum_{o=1}^{N_{\mathrm{O}}} s_{\mathrm{O},o}[n,m] + \sum_{i=1}^{N_{\mathrm{I}}} s_{\mathrm{I},i}[n,m] + \upsilon[n,m] \, ,
\label{eq:signal-model}
\end{equation}
where $s_{\mathrm{O},o}[n,m]$ is the $o^{th}$ object reflection, $s_{\mathrm{I},i}[n,m]$ corresponds to interference from the $i^{th}$ interferer assuming $N_{\mathrm{I}}$ interferers, and $\upsilon[n,m]$ models the noise.

In the simulated radar signal we use \emph{additive white Gaussian noise (AWGN)} to model the receiver noise and point objects with random distances and velocities to model the object reflections. Real measurements already contain object reflections mixed with receiver noise. In both cases, the interference is simulated and added to the time domain object reflections according to Equation~\ref{eq:signal-model}. Noncoherent mutual interference essentially generates time-limited broadband disturbances, see~\cite{TOT18,Kim2018} for details.

State-of-the-art (``classical") interference mitigation methods are mostly signal processing algorithms that are applied either on the time domain signal $s_{\mathrm{IF}}[n,m]$ or on the frequency domain signal $S_{\mathrm{R}}[n,m]$ after the first DFT. The CNN-based method used in this paper, also denoted \emph{Range-Doppler Denoising (RDD)\label{rd-denoising}}, is applied on the RD map after the second DFT.

\section{CNN model architecture}
\label{sec:methodology}

The interference mitigation and denoising CNN architecture is based on the models from \cite{Rock1907:Complex}. \emph{Range Doppler Denoising} (\emph{RDD}, as labeled in Figure \ref{fig:spchain_classical}) is used for evaluation and comparison, because of its superior performance on past experiments with simulated data.

\begin{figure*}
	\centering
	\resizebox {0.9\textwidth} {3.5cm} {
		\input{./cnn_arch.tex}
	}
	\caption{CNN architecture for radar signal denoising. It uses \emph{ReLu}, \emph{Batch Normalization (BN)} and the convolution operation $\textrm{Conv}(i, o, (s_1\times s_2))$, for $i$ input channels, $o$ output channels and a kernel size of $s_{1} \times s_{2}$.}
	\label{fig:cnn_arch}
\end{figure*}
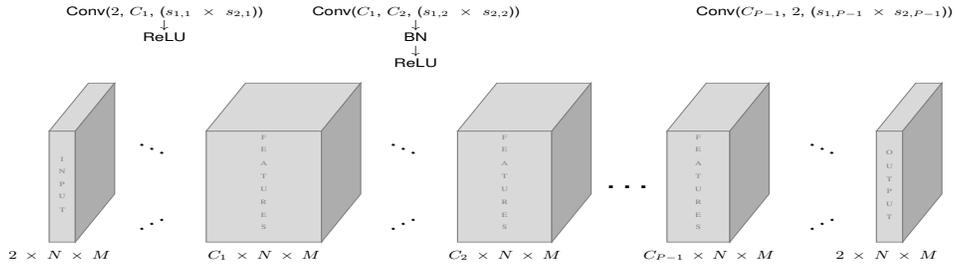

Figure \ref{fig:cnn_arch} illustrates the CNN-based architecture, which consists entirely of convolutional composite layers. The first layer performs convolution operations and \emph{ReLu} \cite{journals/jmlr/GlorotBB11} activation functions; subsequent layers additionally include \emph{Batch Normalization (BN)} \cite{DBLP:journals/corr/IoffeS15}. An exception is the last layer, which uses a linear activation function and two kernels in order to map to the real and imaginary data. The amount of kernels in a layer is chosen to be a power-of-two and decreasing for subsequent layers, e.g. [$2^6, 2^5, 2^4, 2^2$], as inspired by \cite{jiang2019deep}.

RDD is applied to radar snapshots after the second DFT (RD maps), hence the input samples are complex valued patches of size $N \times M$. Square kernels are used in combination with zero-padding, such that the inputs and outputs for each layer have the same spatial dimension. We use two input channels in order to represent the real- and imaginary parts of the complex valued input. For training the network we use the \emph{mean squared error (MSE)} loss function and  the \emph{Adam} \cite{DBLP:journals/corr/KingmaB14} algorithm.

\section{Experimental setup}

\subsection{Data sets}
In our experiments we evaluate two data sets. The first one is purely simulated including object reflections, noise and interference. The second data set consists of real-world measurements, that are combined with simulated interferences. This way we have access to training inputs and their corresponding targets with a limited measurement expenditure. Nevertheless, realistic scenarios are the basis of training and evaluation, and thus give an insight of interference mitigation performance on real-world data. Both data sets are split into three partitions for training ($2500$ snapshots), validation ($250$ snapshots) and testing ($250$ snapshots) the models. Data from a single measurement, consisting of 32 snapshots and sixteen antennas, are exclusively contained either in the training, validation, or test set.

\subsubsection{Simulation}
\label{sec:simulation}
The basic receive IF signal is generated according to~\eqref{eq:signal-model} and processed as described in Section~\ref{sec:sigmod}. The signals are generated based on several parameters, that are sampled from uniform distributions $\mathcal{U}(min,max)$ in the respective domains. Among them are the number of objects $\mathcal{U}(1,20)$ and for each object the relative distance $\mathcal{U}(0\mathrm{m},100\mathrm{m})$ and velocity $\mathcal{U}(-20\mathrm{m/s},20\mathrm{m/s})$.

The interferer parameters are uniformly sampled within the ranges listed in Table \ref{tab:interferer-param}, while the ego radar parameters (see Table~\ref{tab:victim-radar-param}) are constant for all simulations. The \emph{signal-to-noise ratio (SNR)} and the \emph{signal-plus-noise-to-interference ratio (SNIR)} are used to scale the noise and interference powers relative to the object-signal power and object-signal-plus-noise power respectively, when generating the interfered and noisy time domain signal $s_{IF}[n,m]$.

Figure~\ref{fig:sim-rd-interfered-and-target} shows a RD map processed from a simulated signal with five objects, where Figure~\ref{fig:sim-rd-interfered} shows an interfered signal and Figure~\ref{fig:sim-rd-target} shows the corresponding clean data with AWGN.

\begin{table}[ht]
	\centering
	\caption{Ranges of interference and noise parameters.}
	\begin{tabular}{l l r r}
		Parameter & & & Value \\
		\hline
		$f_{\textrm{0,I}}$ & Sweep start frequency & $78.9 \textrm{GHz}$ & $79.1 \textrm{GHz}$ \\
		$B_\textrm{I}$ & Sweep bandwidth & $0.15 \textrm{GHz}$ & $0.25 \textrm{GHz}$ \\
		$T_\textrm{I}$ & Sweep duration & $12 \mu \textrm{s}$  & $24 \mu \textrm{s}$\\
		$\textrm{SNR}$ & Signal-to-noise-ratio & $-15.5 \textrm{dB}$ & $-0.5 \textrm{dB}$ \\
		$\textrm{SNIR}$ & Signal-plus-noise- & $15 \textrm{dB}$ & $35 \textrm{dB}$ \\
		& to-interference-ratio & &  \\
	\end{tabular}
	\label{tab:interferer-param}
	\vspace{-3mm}
\end{table}

\begin{table}[ht]
	\centering
	\caption{Ego radar and signal processing parameters for simulation and measurements.}
	\begin{tabular}{l l r}
		Parameter & & Value \\
		\hline
		$f_{\textrm{0,I}}$ & Sweep start frequency & $79 \textrm{GHz}$ \\
		$B_\textrm{I}$ & Sweep bandwidth & $0.27 \textrm{GHz}$ \\
		$T_\textrm{I}$ & Sweep duration & $12.8 \mu \textrm{s}$ \\
		$B_{\textrm{IF,V}}$ & IF bandwidth & $10 \textrm{MHz}$ \\
		$N$ & Number of fast-time samples & $512$ \\
		$M$ & Number of slow-time samples/ ramps & $128$ \\
		$A$ & Number of antennas & $16$ \\
		$w$ & Window type & $\textrm{Hann}$ \\
	\end{tabular}
	\label{tab:victim-radar-param}
	\vspace{-3mm}
\end{table}

\subsubsection{Real-world measurements}
The measurements were recorded in typical inner-city traffic scenarios. We used a cargo bicycle with a radar apparatus mounted on the front and additional measurement equipment in the cargo container. The device was configured according to the parameters in Table \ref{tab:victim-radar-param}. One measurement denotes 32 consecutive snapshots recorded with sixteen antennas, where each measurement snapshot is associated with a wide-angle camera picture for reference. An input sample for the CNN, i.e. a RD map, is processed from one measurement snapshot of a single antenna.

\begin{figure*}
	\footnotesize
	\centering
	\resizebox {\textwidth} {!} {
	\begin{minipage}{.575\textwidth}
		\centering
		\subfigure[Interfered]{
				\input{./interfered_0_id0.tex}
			\label{fig:sim-rd-interfered}
		}
		\hspace{-4mm}
		\subfigure[Clean]{
				\input{./clean_0_id0.tex}
			\label{fig:sim-rd-target}
		}
		\captionof{figure}{Exemplary RD magnitude spectra of a simulated scenario with five objects in dB.}
		\label{fig:sim-rd-interfered-and-target}
		\label{fig:test2}
	\end{minipage}%
	\begin{minipage}{0.05\textwidth}
		\hspace{1mm}
	\end{minipage}
	\begin{minipage}{0.875\textwidth}
	\centering
	\subfigure[Interfered]{
			\input{./interfered_128_id0.tex}
		\label{fig:meas-rd-interfered}
	}
	\hspace{-4mm}
	\subfigure[Clean]{
			\input{./clean_128_id0.tex}
		\label{fig:meas-rd-target}
	}
	\hspace{-4mm}
	\subfigure[Camera]{
		\raisebox{9.75mm}{\includegraphics[height=3.2cm,width=5.5cm]{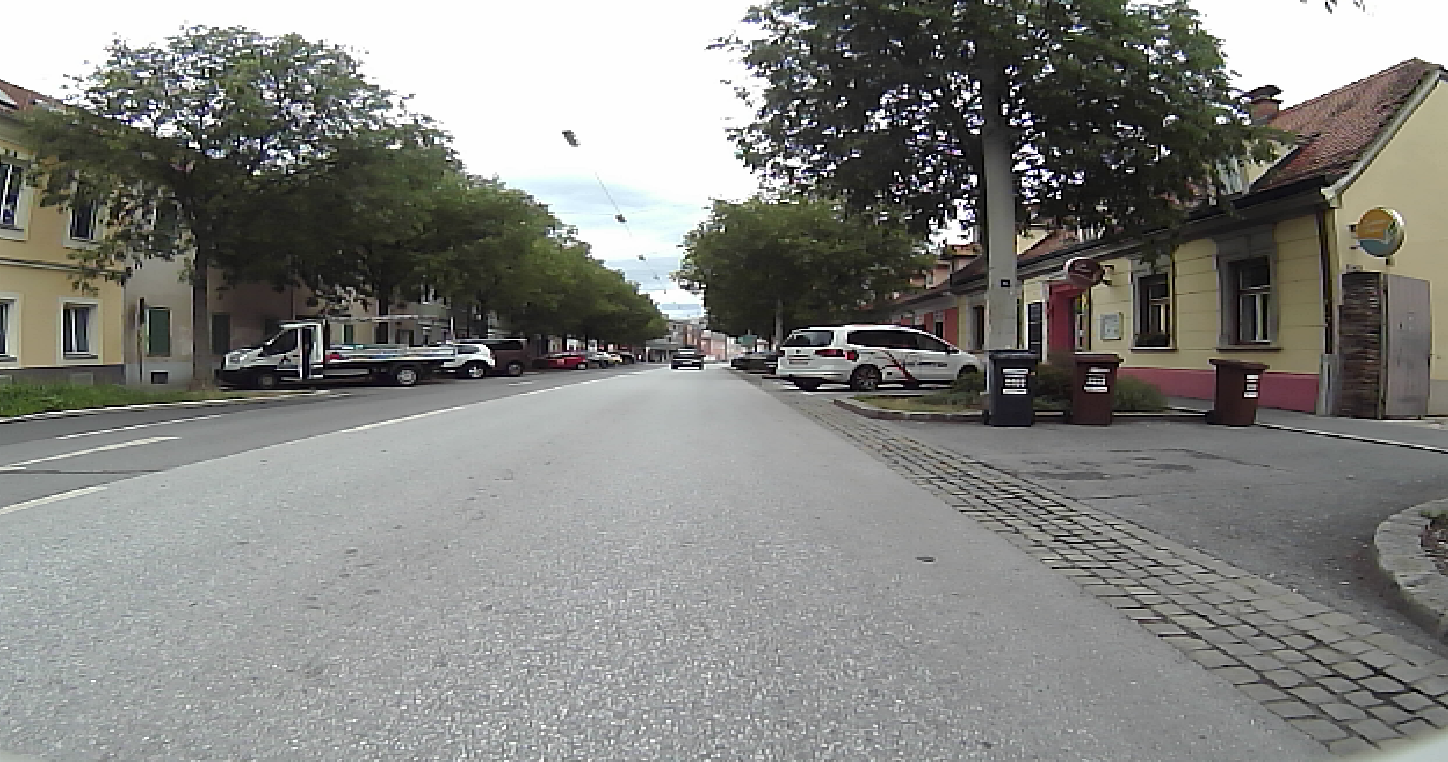}}
		\label{fig:meas-camera}
	}
	\captionof{figure}{Exemplary RD magnitude spectra of a real-world measurement in dB and a reference camera snapshot.}
	\label{fig:meas-rd-interfered-and-target}
\end{minipage}
}
\vspace{-5mm}
\end{figure*}

The measurement signal consists of object reflections (static and moving) as well as receiver noise. The interference is simulated as described in Section~\ref{sec:simulation}. The SNIR is used for scaling the interference relative to the object signal plus receiver noise power. Figure~\ref{fig:meas-rd-interfered-and-target} shows a RD map processed from a real-world measurement, where Figure~\ref{fig:meas-rd-interfered} shows an interfered signal, Figure~\ref{fig:meas-rd-target} shows the corresponding clean signal and Figures~\ref{fig:meas-camera} shows the respective camera snapshot for reference.

\subsubsection{Experimental analysis of simulation and measurements}
The simulated signal is modeled according to reflections from point objects, which results in single, clear and well-shaped object peaks in the RD map. All distances, velocities and angles are randomly sampled, i.e. there is no observable bias towards object peak positions in the RD map.

In the real-world measurements on the other hand, we observe more complex objects, which consist of object peak clusters that are often distributed along the distance as well as the velocity axis of the RD map. Furthermore, a strong bias of object velocities towards the negative velocity of the measurement vehicle is present. This bias results from static objects; they are contained in velocity bins close to the negative ego velocity. Also, there exist strong reflections within the first few meters at a relative velocity of zero, i.e. the reflections of the radar and the measurement vehicle themselves. Another bias, though less severe, can be observed regarding the physical positions of moving objects, caused by the relative position of bicycle lanes to car lanes in the measurement environment.

\subsection{Performance measures}

\subsubsection{Quantitative measures}

The \emph{signal-to-interference-plus-noise ratio (SINR)} is used as performance measure. It is defined as the ratio of signal power at the object peaks to the noise floor~\cite{Rock1907:Complex}. The SINR directly relates to the object detection sensitivity~\cite{performance-comparison-interf-mitigation}, i.e. it significantly influences the chance that an object is detected on the RD map.


The CA-CFAR detector \cite{scharf1991statistical} is used to find the RD map positions of the most prominent object peaks in both the simulated and the measurement data. We apply the detection algorithm with a window of $6 \times 8$ and two guard cells in each dimension. For peaks close to borders, we only consider cells lying within the RD map as reference window cells.

\subsubsection{Qualitative measures}

During visual inspection of the RD map, we consider criteria such as object peak and noise floor magnitude, object peak location, resolution and distortion as well as artifact appearances.

\subsection{Mitigation methods selected for comparison}\label{subsec:setupcomp}

\emph{Zeroing}~\cite{Fischer}, \emph{Iterative method with adaptive thresholding (IMAT)}~\cite{Bechter2017a} and \emph{Ramp filtering}~\cite{WAG18} are chosen as representative state-of-the-art signal processing algorithms for comparison. See \cite{Rock1907:Complex} for an overview of these methods.

Note that zeroing and IMAT require the detection of interfered IF signal samples. In this paper the detection step is assumed to be optimal. In practice however this is not the case, which may have a strong impact on the performance of these algorithms~\cite{performance-comparison-interf-mitigation}.

%
%
%


\section{Experimental results}
First, an extensive parameter search is performed to find suitable CNN-architectures for the simulation and measurement data sets each. The best model architecture is used for finite sample size performance comparison on real-world data. For these evaluations we use the same test set, namely with real measurements, and vary the training set consisting either of the simulated data, real measurements, or both in the context of transfer learning. Finally, we provide a performance comparison with classical interference mitigation methods.

\subsection{Parameter search for a suitable CNN-architecture}
\label{sec:results-arch}
All models are based on the architecture described in Section~\ref{sec:methodology}. In order to find suitable hyper parameters, we run simulations using a different number of layers ($2,3,...,10$) and maximal number of kernels ($2^n, n=3,4,...,8$). A kernel size of $3 \times 3$ is used for all simulations, because of its clear superiority on past experiments.

\begin{figure*}
	\centering
	\subfigure[Parameter]{
		\resizebox {0.3\textwidth} {!} {
			\input{./arch_comparison_parameter_id0.tex}
		}
		\label{fig:arch-num-params}
	}
	\subfigure[Layer]{
		\resizebox {0.3\textwidth} {!} {
			\input{./arch_comparison_layer_id0.tex}
		}
		\label{fig:arch-num-layer}
	}
	\subfigure[Kernels]{
		\resizebox {0.3\textwidth} {!} {
		\input{./arch_comparison_filter_id0.tex}
		}
		\label{fig:arch-num-filter}
	}

	\caption{CNN architecture analysis regarding the performance-model-complexity relation. Table~\ref{tab:arch-comparison-top-results} contains details on selected models marked with the labels A (best on simulation), B (best on measurements), C (smallest on simulation) and D (smallest on measurements).}
	\label{fig:arch}
\end{figure*}
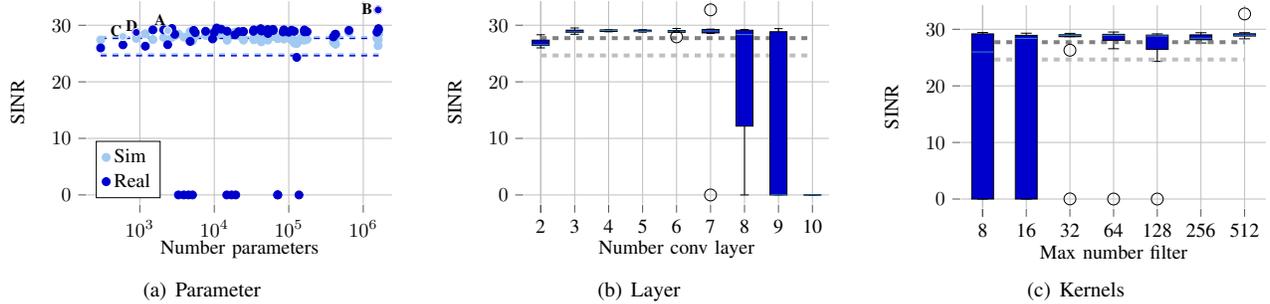

Figure~\ref{fig:arch} shows the SINR based performance comparison of all evaluated architectures. The performance-to-model-complexity relation can be observed in Figure~\ref{fig:arch-num-params}; results for simulated (light blue) and real measurement (dark blue) data are shown with regard to the number of CNN-parameters. The clean and interfered SINRs, denoted $\textrm{SINR}^\textrm{Clean}_\textrm{Sim,Real}$ and $\textrm{SINR}^\textrm{Interfered}_\textrm{Sim,Real}$, are marked with a horizontal line in the corresponding color. The best performing models (A, B) and the smallest models with acceptable performance (C, D) are labeled and respective details are listed in Table~\ref{tab:arch-comparison-top-results}.

The respective best model for each data set reaches outstanding interference mitigation and denoising performance with $\textrm{SINR}^\textrm{A}_\textrm{Sim}=29.03 \textrm{dB}$ and $\textrm{SINR}^\textrm{B}_\textrm{Real}=32.71 \textrm{dB}$. Note, that this metric is significantly higher than the SINR of the 'clean' data without interference, i.e. $\textrm{SINR}^\textrm{Clean}_\textrm{Sim}=27.95 \textrm{dB}$ and $\textrm{SINR}^\textrm{Clean}_\textrm{Real}=27.72 \textrm{dB}$. Thus, the CNN-based models are well suited for the interference mitigation task and have an additional denoising effect. Also, the required model sizes are very small in the context of deep learning. Already models with less than $10^3$ parameters produce excellent results for both data sets. Models for simulated data require even less parameters than models for measurement data in order to reach the 'clean' SINR. The small variance and high values over different network architectures when trained on simulated data indicates more robustness regarding the choice of hyper parameters. Note, that for measurement data the maximal achieved SINR improvement is higher than for simulated data.

Figures ~\ref{fig:arch-num-layer} and ~\ref{fig:arch-num-filter} analyze the performance on measurement data according to the amount of CNN layers and convolutional kernels respectively. A high mean and low variance without negative outliers per hyper parameter indicate good performing and robust CNN model architectures. Thus, models with $3$ to $9$ layers and $16$ to $512$ kernels per layer indicates a robust training process and denoising performance. We choose an architecture with $4$ layers and $[512,32,16,2]$ kernels in these layers for further evaluations.

\begin{table}
	\centering
	\caption{Details for selected CNN-architectures.}
	\begin{tabular}{l l l r r >{\bfseries}r >{\bfseries}r}
		Marker & Note & Data  & Layers & Kernels & Parameters & SINR \\
		\hline
		A & Best & Sim & $2$ & $64$ & 2370 & 29.03 dB \\
		B & Best & Real & $7$ & $512$ & 1582834 & 32.71 dB \\
		C & Small & Sim & $2$ & $16$ & 594 & 28.02 dB \\
		D & Small & Real & $3$ & $8$ & 898 & 28.73 dB \\
	\end{tabular}
	\label{tab:arch-comparison-top-results}
	\vspace{-3mm}
\end{table}

\subsection{Finite sample size performance comparison on real-world data}
We investigate the interference mitigation and denoising capabilities on real measurements dependent on the amount of training samples. Therefore the same test set, namely with measurement data, is used for all evaluations. For training we consider three different data sets consisting of: simulated data (Sim), measurement data (Real), and both, simulated data to pre-train the model and measurement data for fine-tuning, in the context of transfer learning (Transfer). In the case of simulated training data, we run simulations using also simulated data for validation (Sim), and using measurement data for validation (Sim+VReal). For each data set we train the model using ${50, 100,..., 600}$ samples. In the case of transfer learning, we pre-train the model with $500$ simulated samples and fine-tune with the respective amount of measurements.

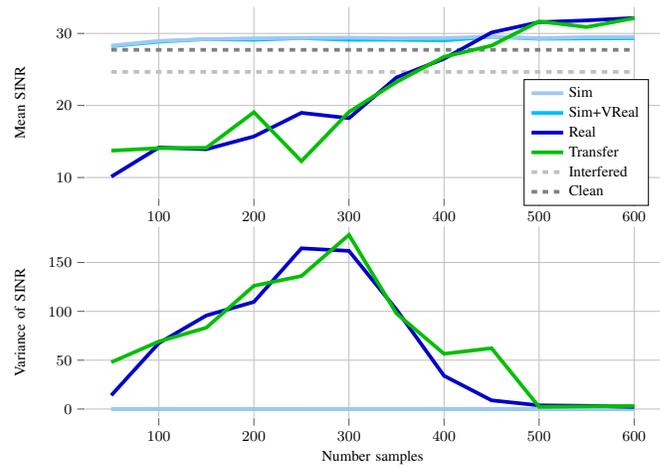
\begin{figure}[ht]
	\footnotesize
	\centering
		\resizebox {\columnwidth} {!} {
			\input{./arch_comparison_samples_mean_id0.tex}
		}
		\resizebox {\columnwidth} {!} {
		\input{./arch_comparison_samples_var_id0.tex}
		}
	\caption{Relation of SINR to the number of samples in the training set.}
	\label{fig:num_samples}
\end{figure}

Figure~\ref{fig:num_samples} shows the relation of SINR performance to the number of samples in the training set. The top and bottom figures show the mean and variance of the SINR, based on twenty simulations per configuration using a randomly selected training subset.

According to our evaluations, also the training with simulated data results in a very high interference mitigation and denoising performance. This indicates, that our model indeed learns to remove the interference and noise instead of learning the object scenarios. Naturally, the outstanding performance is possible, because we use simulated interference also for testing. Nonetheless, the evaluation shows, that interference mitigation can be generalized to unseen object scenarios, such as real-world measurements, as long as realistic interference is used during training.

Training on simulated data seems to be very stable according to the small variance over all training set sizes. This may result from the simpler nature of simulated data, which is beneficial for the learning process. Whether simulated (Sim) or real (Sim+VReal) data is used for validation does not seem to have a strong influence on the performance.

For training with measurement data and transfer learning, the SINR is reduced for fewer training samples; it increases with a rising number of samples until it reaches $\textrm{SINR}^\textrm{Clean}_\textrm{Real}$ with around 400 samples and surpasses the performance of models trained on simulated data with 450 (Real) or 500 (Transfer) samples. During this rise of mean SINR, the variance increases as the model's training progress highly depends on the significance of the randomly sampled training subset. With more than 300 samples, the variance drops again and stabilizes at a low level with more than 500 samples. Advantages of transfer learning over training with measurement data can only be observed for a very small amount of training samples (50 samples) and thus is not considered beneficial for our experiments.

\subsection{Performance comparison with classical interference mitigation methods}

\begin{figure}[tb]
	\footnotesize
	\centering
	\resizebox {\columnwidth} {!} {
	\subfigure[Interfered]{
		\input{./eval_evaluation_test_doppler-range_matrix_interfered_p19_id0.tex}
		\label{fig:exemplary_rd_interfered}
	}
	\hspace{-4mm}
	\subfigure[Clean]{
		\input{./eval_evaluation_test_doppler-range_matrix_targets_p19_id0.tex}
		\label{fig:exemplary_rd_clean}
	}
	\hspace{-4mm}
	\subfigure[After CNN Mitigation]{
		\input{./eval_evaluation_test_doppler-range_matrix_predictions_p19_id0.tex}
		\label{fig:exemplary_rd_cnn}
	}
	}
	\caption{Exemplary RD magnitude spectra from the measurements test set.}
	\label{fig:exemplary_rd}
	\vspace{-5mm}
\end{figure}
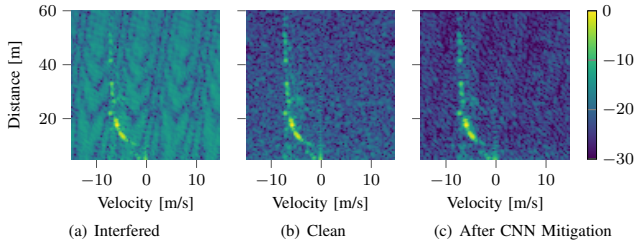

Three classical interference mitigation methods, as described in Section \ref{subsec:setupcomp}, were implemented and evaluated using SINR. The results are statistically compared to the CNN-based model, that was trained and validated on $2500$ and $250$ measurement snapshots, respectively. For the evaluation we used a Monte-Carlo-Simulation with $250$ measurement snapshots. Figure~\ref{fig:cdf-classical} shows the empirical \emph{cumulative density function (CDF)} of their evaluated SINR values. The 'clean' measurement and interfered signals are included as reference. One of the measurement snapshot RD maps with interference, without interference and with mitigation using the CNN-based model is displayed in Figure \ref{fig:exemplary_rd}.

The three classical methods, i.e. zeroing, IMAT and Ramp filtering, all improve the SINR slightly in measurement snapshots with strong interference. For moderate to weak interference on the other hand, they are not capable of removing interference effects and even decrease the SINR of the interfered signal, i.e. perform worse than without mitigation. The CNN-based model outperforms the classical methods for all tested interference levels and even surpasses the SINR of the 'clean' signal without interference, thus it performs additional denoising of the noisy real-world measurements. The shape of the CDF indicates, that this outstanding interference mitigation and denoising performance is also robust over all test set samples.

\begin{figure}[tb]
	\footnotesize
	\centering
	\resizebox {0.95\columnwidth} {!} {
		\input{./cdf_classical.tex}
	}
	\caption{CDF comparison of SINR with classical methods.}
	\label{fig:cdf-classical}
\end{figure}
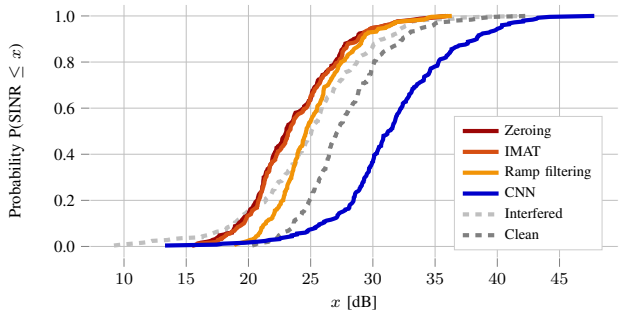

\section{Conclusion}

In this paper, we use CNN-based mutual interference mitigation and denoising models on real-world data. The training and evaluation framework was extended to handle real measurement data and simulated interference was used to obtain input- and output-pairs for training and evaluating the CNN models.

Our experiments show, that the CNN-based interference mitigation approach is also applicable to real-world measurements and results in an outstanding and robust performance. Noteworthy is the small model size, i.e. starting with around $10^3$ parameters, that is required for this task. This is crucial for the use in real-time embedded systems. We illustrate the impact of training set coverage to the performance and robustness of the models. The use of simulated data for training can reduce the amount of required real measurements.
In a performance comparison with classical interference mitigation methods, the CNN-based model outperforms the state of the art and shows robust behavior in our Monte-Carlo-Simulation.

The most important task in the future is to collect real interference measurements and to evaluate the generalization capabilities of the CNN-based models on these data.

\section*{Acknowledgments}
This work was supported by the Austrian Research Promotion Agency (FFG) under the project SAHaRA (17774193) and NVIDIA by providing GPUs.

\bibliographystyle{ieeetr}
\bibliography{bibliography}

\end{document}

%% file: spchain_classical.tex
\tikzstyle{block}=[draw, fill=black!10, text width=5em, text centered, minimum height=6em]
\tikzstyle{blockfocus}=[block, draw=red, thick, dashed,rounded corners]
\tikzstyle{blockinactive}=[block, opacity=.3]

\tikzstyle{arrow}=[very thick, ->]
\tikzstyle{arrowinactive}=[arrow, opacity=.2]

\begin{tikzpicture}[scale=0.7, transform shape]
\begin{scope}[node distance=1cm and 2.5cm]

\node[block] (dft2) at (0,0) {DFT over $m$ \\ for each $n$};
\node[block, above left = of dft2] (rs) {Radar Sensor};
\node[block, above = of dft2] (tdp) {Time Domain Denoising (TDD)};
\node[block, above right = of dft2] (dft1) {DFT over $n$ \\ for each $m$};
\node[block, left = of dft2] (rdd) {Range-Doppler Denoising (RDD)};
\node[block, right = of dft2] (rpd) {Range-Profile Denoising (RPD)};
\node[block, below left = of dft2] (od) {Object Detection};
\node[block, below = of dft2] (ae) {Angle Estimation};
\node[block, below right = of dft2] (fp) {Further Processing};

\draw[arrow] (rs) -- (tdp) node[midway, above] () {$s_{\mathrm{IF}}[n,m]$} node [midway,below] () {${(N{\times}M)}$};
\draw[arrow] (tdp) -- (dft1) node[midway, above] () {$\tilde{s}_{\mathrm{IF}}[n,m]$};
\draw[arrow] (dft1) -- (rpd) node[pos=0.3, left] () {$S_{\mathrm{R}}[n,m]$};
\draw[arrow] (rpd) -- (dft2) node[midway, above] () {$\tilde{S}_{\mathrm{R}}[n,m]$};
\draw[arrow] (dft2) -- (rdd) node[midway, above] () {$S_{\mathrm{RD}}[n,m]$};
\draw[arrow] (rdd) -- (od) node[pos=0.7, right] () {$\tilde{S}_{\mathrm{RD}}[n,m]$};
\draw[arrow] (od) -- (ae) node[midway, above] () {object peaks};
\draw[arrow] (ae) -- (fp) node[midway, above] () {objects};

\draw[red,thick,dashed,rounded corners] ($(rpd.north west)+(-0.4,0.4)$)  rectangle ($(rpd.south east)+(0.4,-0.4)$);
\draw[red,thick,dashed,rounded corners] ($(rdd.north west)+(-0.4,0.4)$)  rectangle ($(rdd.south east)+(0.4,-0.4)$);

\draw[blue,thick,dashed,rounded corners] ($(tdp.north west)+(-0.2,0.2)$)  rectangle ($(tdp.south east)+(0.2,-0.2)$);
\draw[blue,thick,dashed,rounded corners] ($(rpd.north west)+(-0.2,0.2)$)  rectangle ($(rpd.south east)+(0.2,-0.2)$);

\end{scope}
\end{tikzpicture}

%% file: cnn_arch.tex
\tikzset{pics/fake box/.style args={
#1 with dimensions #2 and #3 and #4 and text #5 at ypos #6}{
code={
\draw[gray,ultra thin,fill=#1]  (0,0,0) coordinate(-front-bottom-left) to
++ (0,#3,0) coordinate(-front-top-right) --++
(#2,0,0) coordinate(-front-top-right) --++ (0,-#3,0) 
coordinate(-front-bottom-right) -- cycle;
\draw[gray,ultra thin,fill=#1] (0,#3,0)  --++ 
 (0,0,#4) coordinate(-back-top-left) --++ (#2,0,0) 
 coordinate(-back-top-right) --++ (0,0,-#4)  -- cycle;
\draw[gray,ultra thin,fill=#1!80!black] (#2,0,0) --++ (0,0,#4) coordinate(-back-bottom-right)
--++ (0,#3,0) --++ (0,0,-#4) -- cycle;
\path[gray,decorate,decoration={text effects along path,text align=fit to path,text={#5}},font=\tiny] (#2/2,#6,0) -- (#2/2,0,0);
}
}}
\tikzset{circle dotted/.style={dash pattern=on .05mm off 2mm,
                                         line cap=round}}

\begin{tikzpicture}[x={(1,0)},y={(0,1)},z={({cos(60)},{sin(60)})},
font=\sffamily\small,scale=2]
%


\node (box2-end-node) at (2.5,0,0) {};
\node (box3-node) at (3.5,0,0) {};

\draw pic (box1) at (-2.4,-1.5/2,0) {fake box=white!70!gray with dimensions 0.5 and 3 and 1.5 and text INPUT at ypos 2.4};
\draw pic (box1) at (-0.9,-1.5/2,0) {fake box=white!70!gray with dimensions 2.2 and 3 and 1.5 and text FEATURES at ypos 3.0};
\draw pic (box2) at (1.5,-1.5/2,0) {fake box=white!70!gray with dimensions 1.8 and 3 and 1.5 and text FEATURES at ypos 3.0};
\draw pic (box3) at (3.5,-1.5/2,0) {fake box=white!70!gray with dimensions 1.2 and 3 and 1.5 and text FEATURES at ypos 3.0};
\draw pic (box4) at (5.5,-1.5/2,0) {fake box=white!70!gray with dimensions 0.5 and 3 and 1.5 and text OUTPUT at ypos 2.6};

\node[text width=10cm,align=center,anchor=north] at (-1.3,2.5,0) {Conv($2$, $C_1$, ($s_{1,1}\!\times\! s_{2,1}$))\\$\downarrow$\\ReLU};

\node[text width=10cm,align=center,anchor=north] at (1.1,2.5,0) {Conv($C_1$, $C_2$, ($s_{1,2}\!\times\! s_{2,2}$))\\$\downarrow$\\BN\\$\downarrow$\\ReLU};

\node[text width=10cm,align=center,anchor=north] at (5,2.5,0) {Conv($C_{P-1}$, $2$, ($s_{1,P-1}\!\times\! s_{2,P-1}$))};

\path (-1.9,0.5,0) -- node[auto=false,rotate=-30]{\Large\ldots} (-0.9,0.5,0);
\path (-1.9,-0.5,0) -- node[auto=false,rotate=30]{\Large\ldots} (-0.9,-0.5,0);

\path (0.5,0.5,0) -- node[auto=false,rotate=-30]{\Large\ldots} (1.5,0.5,0);
\path (0.5,-0.5,0) -- node[auto=false,rotate=30]{\Large\ldots} (1.5,-0.5,0);

\path (2.8,0,0) -- node[auto=false]{\huge\ldots} (3.5,0,0);

\path (4.5,0.5,0) -- node[auto=false,rotate=-30]{\Large\ldots} (5.5,0.5,0);
\path (4.5,-0.5,0) -- node[auto=false,rotate=30]{\Large\ldots} (5.5,-0.5,0);

\node[text width=5cm,align=center,anchor=north] at (-2.3,-0.8,0) {$2 \times N \times M$};

\node[text width=5cm,align=center,anchor=north] at (-0.35,-0.8,0) {$C_1 \times N \times M$};

\node[text width=5cm,align=center,anchor=north] at (1.95,-0.8,0) {$C_2 \times N \times M$};

\node[text width=5cm,align=center,anchor=north] at (3.9,-0.8,0) {$C_{P-1} \times N \times M$};

\node[text width=5cm,align=center,anchor=north] at (5.6,-0.8,0) {$2 \times N \times M$};

%
\end{tikzpicture}

%% file: interfered_0_id0.tex
\begin{tikzpicture}

\begin{axis}[
scale only axis=true,
width=3.2cm,
height=3.2cm,
axis background/.style={fill=white!89.80392156862746!black},
axis line style={white},
point meta max=0,
point meta min=-30,
tick align=outside,
tick pos=left,
x grid style={white},
xlabel={Velocity [m/s]},
xmajorgrids,
xmin=-15, xmax=14.6703296703297,
xtick style={color=white!33.33333333333333!black},
y grid style={white},
ylabel={Distance [m]},
y label style={at={(axis description cs:0.2,.5)},anchor=south},
ymajorgrids,
ymin=5, ymax=60,
ytick style={color=white!33.33333333333333!black}
]
\addplot graphics [includegraphics cmd=\pgfimage,xmin=-15, xmax=14.6703296703297, ymin=5, ymax=60] {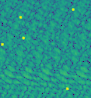};
\end{axis}

\end{tikzpicture}

%% file: clean_0_id0.tex
\begin{tikzpicture}

\begin{axis}[
scale only axis=true,
width=3.2cm,
height=3.2cm,
axis background/.style={fill=white!89.80392156862746!black},
axis line style={white},
colorbar,
colorbar style={ylabel={}},
colorbar/width=1.5mm,
colormap/viridis,
point meta max=0,
point meta min=-30,
tick align=outside,
tick pos=left,
x grid style={white},
xlabel={Velocity [m/s]},
xmajorgrids,
xmin=-15, xmax=14.6703296703297,
xtick style={color=white!33.33333333333333!black},
y grid style={white},
yticklabels={,,},
ymajorgrids,
ymin=5, ymax=60,
ytick style={color=white!33.33333333333333!black}
]
\addplot graphics [includegraphics cmd=\pgfimage,xmin=-15, xmax=14.6703296703297, ymin=5, ymax=60] {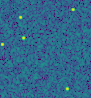};
\end{axis}

\end{tikzpicture}

%% file: interfered_128_id0.tex
\begin{tikzpicture}

\begin{axis}[
scale only axis=true,
width=3.2cm,
height=3.2cm,
axis background/.style={fill=white!89.80392156862746!black},
axis line style={white},
point meta max=0,
point meta min=-50,
tick align=outside,
tick pos=left,
x grid style={white},
xlabel={Velocity [m/s]},
xmajorgrids,
xmin=-14.9237578125, xmax=14.9237578125,
y grid style={white},
ylabel={Distance [m]},
y label style={at={(axis description cs:0.2,.5)},anchor=south},
ymajorgrids,
ymin=0, ymax=60
]
\addplot graphics [includegraphics cmd=\pgfimage,xmin=-14.9237578125, xmax=14.9237578125, ymin=0, ymax=60] {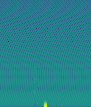};

\end{axis}
\end{tikzpicture}

%% file: clean_128_id0.tex
\begin{tikzpicture}

\begin{axis}[
scale only axis=true,
width=3.2cm,
height=3.2cm,
axis background/.style={fill=white!89.80392156862746!black},
axis line style={white},
colorbar,
colorbar style={ylabel={}},
colorbar/width=1.5mm,
colormap/viridis,
point meta max=0,
point meta min=-50,
tick align=outside,
tick pos=left,
x grid style={white},
xlabel={Velocity [m/s]},
xmajorgrids,
xmin=-14.9237578125, xmax=14.9237578125,
y grid style={white},
yticklabels={,,},
ymajorgrids,
ymin=0, ymax=60
]
\addplot graphics [includegraphics cmd=\pgfimage,xmin=-14.9237578125, xmax=14.9237578125, ymin=0, ymax=60] {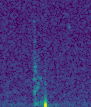};

\end{axis}
\end{tikzpicture}

%% file: arch_comparison_parameter_id0.tex
\begin{tikzpicture}

\definecolor{color0}{rgb}{0.63, 0.79, 0.95}

\begin{axis}[
scale only axis=true,
width=0.3\textwidth,
height=0.2\textwidth,
axis line style={white},
legend cell align={left},
legend style={at={(0.03,0.03)}, anchor=south west}, 
log basis x={10},
tick align=outside,
tick pos=left,
xlabel={Number parameters},
xmajorgrids,
xmin=193.951836016327, xmax=2460542.50505479,
xmode=log,
xtick style={color=white!33.33333333333333!black},
xtick={10,100,1000,10000,100000,1000000,10000000,100000000},
xticklabels={\(\displaystyle {10^{1}}\),\(\displaystyle {10^{2}}\),\(\displaystyle {10^{3}}\),\(\displaystyle {10^{4}}\),\(\displaystyle {10^{5}}\),\(\displaystyle {10^{6}}\),\(\displaystyle {10^{7}}\),\(\displaystyle {10^{8}}\)},
ylabel={SINR},
ymajorgrids,
ymin=-1.64109188401701, ymax=34.3557867072143,
ytick style={color=white!33.33333333333333!black}
]

\addplot [only marks, draw=color0, fill=color0, colormap/viridis]
table{%
x                      y
5554 28.1362620519117
10770 28.2527416960031
898 27.667743290122
42066 28.2597054618259
2946 28.2287503332802
83794 28.4092542394209
298 27.3999970248767
21202 28.17226908852
2370 29.0306280118301
594 28.0230816166463
5098 27.2269808556328
18946 28.7639798492569
61426 28.4457160489803
4738 28.8779467962493
137074 27.2376251955171
425490 26.6444831719272
17058 28.4892933299829
4498 27.7057796508137
71250 27.6520520883294
1601458 28.9630188957337
70738 27.6968260785976
19410 28.7988260631719
146386 27.786173255412
9474 28.7283834344292
80562 27.9559525398577
434802 28.093395763511
641842 28.725696662952
162226 27.3557653759313
1610770 28.3937196538893
1527346 27.6817279925392
1592146 27.4613398521953
1582834 26.390423138014
1498 27.4765260617674
2698 27.8542672054991
14866 27.3532696288517
3898 27.9124650831481
24178 27.6349388321128
10002 27.9081907612393
5298 27.890847538424
33490 27.6437402633758
7650 28.2076212274724
24690 27.3664963683511
52114 27.5735090133255
14706 28.9749118709453
2098 27.8430919327827
12354 28.7068713081064
42802 27.0806974954624
3298 28.4521945188542
162066 27.7050665123713
118450 26.8490009952363
175986 27.5129128535979
127762 26.7570814578568
406866 27.3987974626474
416178 27.0162210324949
397554 26.9975786730847
1186 28.496995553954
52626 27.3787225437566
34002 27.4290606227881
43314 27.3421614152875
85906 27.5653708436552
61938 28.0444819439486
99826 27.7312491606728
109138 27.4362915691829
};
\addlegendentry{Sim}

\addplot [only marks, draw=blue!80.3921568627451!black, fill=blue!80.3921568627451!black, colormap/viridis]
table{%
	x                      y
	34002 29.1780286753807
	43314 28.9951511737308
	52626 28.5513224152134
	99826 28.7753174557709
	24690 28.8819921584142
	109138 29.1204141020361
	127762 24.3210761661598
	162066 29.1692002735686
	118450 28.9286277223312
	175986 28.8619677605596
	85906 29.1883591341982
	416178 28.3492219821834
	61938 29.0855538091763
	397554 27.9426338564964
	162226 28.8405442126035
	641842 29.0965548541938
	2098 29.1973827290819
	1498 29.1852998498678
	2698 29.4276590833633
	14706 -7.13602325249433e-16
	3298 -5.5159531086848e-16
	3898 -8.77537994563491e-16
	1527346 28.7740465316751
	5298 29.0842871695968
	10002 28.7238084542899
	14866 28.9344585733443
	7650 28.9320034598554
	12354 29.3079324478299
	42802 29.0393777317087
	33490 28.9303520294123
	24178 28.8915092636133
	52114 29.244003361846
	1582834 32.7146948231973
	1592146 29.1414001628492
	2946 28.3985515113048
	898 28.7293718986757
	5554 28.6957275512446
	21202 28.859101362274
	10770 29.5150946335932
	594 26.5362980524909
	298 25.9940390959043
	9474 27.5800118017163
	4738 27.139795714837
	83794 28.8167432092145
	1186 26.3005577890913
	42066 29.3939849094197
	71250 2.72310551268741e-05
	2370 26.5623876968039
	18946 28.3159371967028
	4498 9.25754367891156e-16
	17058 -1.15719295986394e-16
	61426 29.2756456636568
	5098 -4.24304085283448e-17
	19410 -3.85730986621315e-16
	137074 9.27683022824262e-16
	1601458 29.4011286995754
	425490 28.4257905511872
	70738 0.0257665784768379
};
\addlegendentry{Real}

\addplot [dashed, semithick, blue!80.3921568627451!black, line width=1pt]
table {%
298 27.7211477013
594 27.7211477013
898 27.7211477013
1186 27.7211477013
1498 27.7211477013
2098 27.7211477013
2370 27.7211477013
2698 27.7211477013
2946 27.7211477013
3298 27.7211477013
3898 27.7211477013
4498 27.7211477013
4738 27.7211477013
5098 27.7211477013
5298 27.7211477013
5554 27.7211477013
7650 27.7211477013
9474 27.7211477013
10002 27.7211477013
10770 27.7211477013
12354 27.7211477013
14706 27.7211477013
14866 27.7211477013
17058 27.7211477013
18946 27.7211477013
19410 27.7211477013
21202 27.7211477013
24178 27.7211477013
24690 27.7211477013
33490 27.7211477013
34002 27.7211477013
42066 27.7211477013
42802 27.7211477013
43314 27.7211477013
52114 27.7211477013
52626 27.7211477013
61426 27.7211477013
61938 27.7211477013
70738 27.7211477013
71250 27.7211477013
83794 27.7211477013
85906 27.7211477013
99826 27.7211477013
109138 27.7211477013
118450 27.7211477013
127762 27.7211477013
137074 27.7211477013
162066 27.7211477013
162226 27.7211477013
175986 27.7211477013
397554 27.7211477013
416178 27.7211477013
425490 27.7211477013
641842 27.7211477013
1527346 27.7211477013
1582834 27.7211477013
1592146 27.7211477013
1601458 27.7211477013
};
\addplot [dashed, semithick, blue!80.3921568627451!black, line width=1pt]
table {%
298 24.6550554292
594 24.6550554292
898 24.6550554292
1186 24.6550554292
1498 24.6550554292
2098 24.6550554292
2370 24.6550554292
2698 24.6550554292
2946 24.6550554292
3298 24.6550554292
3898 24.6550554292
4498 24.6550554292
4738 24.6550554292
5098 24.6550554292
5298 24.6550554292
5554 24.6550554292
7650 24.6550554292
9474 24.6550554292
10002 24.6550554292
10770 24.6550554292
12354 24.6550554292
14706 24.6550554292
14866 24.6550554292
17058 24.6550554292
18946 24.6550554292
19410 24.6550554292
21202 24.6550554292
24178 24.6550554292
24690 24.6550554292
33490 24.6550554292
34002 24.6550554292
42066 24.6550554292
42802 24.6550554292
43314 24.6550554292
52114 24.6550554292
52626 24.6550554292
61426 24.6550554292
61938 24.6550554292
70738 24.6550554292
71250 24.6550554292
83794 24.6550554292
85906 24.6550554292
99826 24.6550554292
109138 24.6550554292
118450 24.6550554292
127762 24.6550554292
137074 24.6550554292
162066 24.6550554292
162226 24.6550554292
175986 24.6550554292
397554 24.6550554292
416178 24.6550554292
425490 24.6550554292
641842 24.6550554292
1527346 24.6550554292
1582834 24.6550554292
1592146 24.6550554292
1601458 24.6550554292
};
\addplot [dashed, semithick, color0, line width=1pt]
table {%
298 27.9480225945
594 27.9480225945
898 27.9480225945
1186 27.9480225945
1498 27.9480225945
2098 27.9480225945
2370 27.9480225945
2698 27.9480225945
2946 27.9480225945
3298 27.9480225945
3898 27.9480225945
4498 27.9480225945
4738 27.9480225945
5098 27.9480225945
5298 27.9480225945
5554 27.9480225945
7650 27.9480225945
9474 27.9480225945
10002 27.9480225945
10770 27.9480225945
12354 27.9480225945
14706 27.9480225945
14866 27.9480225945
17058 27.9480225945
18946 27.9480225945
19410 27.9480225945
21202 27.9480225945
24178 27.9480225945
24690 27.9480225945
33490 27.9480225945
34002 27.9480225945
42066 27.9480225945
42802 27.9480225945
43314 27.9480225945
52114 27.9480225945
52626 27.9480225945
61426 27.9480225945
61938 27.9480225945
70738 27.9480225945
71250 27.9480225945
80562 27.9480225945
83794 27.9480225945
85906 27.9480225945
99826 27.9480225945
109138 27.9480225945
118450 27.9480225945
127762 27.9480225945
137074 27.9480225945
146386 27.9480225945
162066 27.9480225945
162226 27.9480225945
175986 27.9480225945
397554 27.9480225945
406866 27.9480225945
416178 27.9480225945
425490 27.9480225945
434802 27.9480225945
641842 27.9480225945
1527346 27.9480225945
1582834 27.9480225945
1592146 27.9480225945
1601458 27.9480225945
1610770 27.9480225945
};
\addplot [dashed, semithick, color0, line width=1pt]
table {%
298 24.9098443897
594 24.9098443897
898 24.9098443897
1186 24.9098443897
1498 24.9098443897
2098 24.9098443897
2370 24.9098443897
2698 24.9098443897
2946 24.9098443897
3298 24.9098443897
3898 24.9098443897
4498 24.9098443897
4738 24.9098443897
5098 24.9098443897
5298 24.9098443897
5554 24.9098443897
7650 24.9098443897
9474 24.9098443897
10002 24.9098443897
10770 24.9098443897
12354 24.9098443897
14706 24.9098443897
14866 24.9098443897
17058 24.9098443897
18946 24.9098443897
19410 24.9098443897
21202 24.9098443897
24178 24.9098443897
24690 24.9098443897
33490 24.9098443897
34002 24.9098443897
42066 24.9098443897
42802 24.9098443897
43314 24.9098443897
52114 24.9098443897
52626 24.9098443897
61426 24.9098443897
61938 24.9098443897
70738 24.9098443897
71250 24.9098443897
80562 24.9098443897
83794 24.9098443897
85906 24.9098443897
99826 24.9098443897
109138 24.9098443897
118450 24.9098443897
127762 24.9098443897
137074 24.9098443897
146386 24.9098443897
162066 24.9098443897
162226 24.9098443897
175986 24.9098443897
397554 24.9098443897
406866 24.9098443897
416178 24.9098443897
425490 24.9098443897
434802 24.9098443897
641842 24.9098443897
1527346 24.9098443897
1582834 24.9098443897
1592146 24.9098443897
1601458 24.9098443897
1610770 24.9098443897
};

\addplot [only marks, draw=white, fill=color0, colormap/viridis]
table{%
	x                      y
	2370 29.0306280118301
	594 28.0230816166463
};

\addplot [only marks, draw=white, fill=blue!80.3921568627451!black, colormap/viridis]
table{%
	x                      y
	1582834 32.7146948231973
	898 28.7293718986757
};

\draw[] (axis cs:1582834,32.71) -- (axis cs:1582834,32.71);
\node at (axis cs:1582800,33)[
scale=0.8,
anchor=east,
align=center,
text=black,
rotate=0.0
]{\bfseries B};
\draw[] (axis cs:898,28.73) -- (axis cs:898,28.73);
\node at (axis cs:1100,30)[
scale=0.8,
anchor=east,
align=center,
text=black,
rotate=0.0
]{\bfseries D};
\draw[] (axis cs:2370,29.03) -- (axis cs:2370,29.03);
\node at (axis cs:2670,31)[
scale=0.8,
anchor=east,
align=center,
text=black,
rotate=0.0
]{\bfseries A};
\draw[] (axis cs:594,28.02) -- (axis cs:594,28.02);
\node at (axis cs:680,29)[
scale=0.8,
anchor=east,
align=center,
text=black,
rotate=0.0
]{\bfseries C};
\end{axis}

\end{tikzpicture}

%% file: arch_comparison_layer_id0.tex
\begin{tikzpicture}

\definecolor{color0}{rgb}{0.203921568627451,0.541176470588235,0.741176470588235}
\definecolor{color1}{rgb}{0,0.749019607843137,1}

\begin{axis}[
scale only axis=true,
width=0.3\textwidth,
height=0.2\textwidth,
axis line style={white},
tick align=outside,
tick pos=left,
xlabel={Number conv layer},
xmajorgrids,
xmin=0.5, xmax=9.5,
xtick style={color=white!33.33333333333333!black},
xtick={1,2,3,4,5,6,7,8,9,1},
xticklabels={2,3,4,5,6,7,8,9,10,},
ylabel={SINR},
ymajorgrids,
ymin=-1.63573474115987, ymax=34.3504295643572,
ytick style={color=white!33.33333333333333!black}
]
\addplot [black]
table {%
1 26.4184279207911
1 25.9940390959043
};
\addplot [black]
table {%
1 27.3599037582767
1 28.3159371967028
};
\addplot [black]
table {%
0.875 25.9940390959043
1.125 25.9940390959043
};
\addplot [black]
table {%
0.875 28.3159371967028
1.125 28.3159371967028
};
\addplot [black]
table {%
2 28.7125497249602
2 28.3985515113048
};
\addplot [black]
table {%
2 29.1265431358468
2 29.5150946335932
};
\addplot [black]
table {%
1.875 28.3985515113048
2.125 28.3985515113048
};
\addplot [black]
table {%
1.875 29.5150946335932
2.125 29.5150946335932
};
\addplot [black]
table {%
3 28.9082253658792
3 28.8405442126035
};
\addplot [black]
table {%
3 29.1772500617182
3 29.1883591341982
};
\addplot [black]
table {%
2.875 28.8405442126035
3.125 28.8405442126035
};
\addplot [black]
table {%
2.875 29.1883591341982
3.125 29.1883591341982
};
\addplot [black]
table {%
4 28.8767385120864
4 28.7753174557709
};
\addplot [black]
table {%
4 29.1372917647872
4 29.1973827290819
};
\addplot [black]
table {%
3.875 28.7753174557709
4.125 28.7753174557709
};
\addplot [black]
table {%
3.875 29.1973827290819
4.125 29.1973827290819
};
\addplot [black]
table {%
5 28.7489274929825
5 28.7238084542899
};
\addplot [black]
table {%
5 29.0577826378834
5 29.4276590833633
};
\addplot [black]
table {%
4.875 28.7238084542899
5.125 28.7238084542899
};
\addplot [black]
table {%
4.875 29.4276590833633
5.125 29.4276590833633
};
\addplot [black, mark=*, mark size=3, mark options={solid,fill opacity=0}, only marks]
table {%
5 27.9426338564964
};
\addplot [black]
table {%
6 28.6456487419929
6 28.5513224152134
};
\addplot [black]
table {%
6 29.2407937687996
6 29.3079324478299
};
\addplot [black]
table {%
5.875 28.5513224152134
6.125 28.5513224152134
};
\addplot [black]
table {%
5.875 29.3079324478299
6.125 29.3079324478299
};
\addplot [black, mark=*, mark size=3, mark options={solid,fill opacity=0}, only marks]
table {%
6 -5.5159531086848e-16
6 32.7146948231973
};
\addplot [black]
table {%
7 12.1605380830799
7 -8.77537994563491e-16
};
\addplot [black]
table {%
7 29.1134769860127
7 29.244003361846
};
\addplot [black]
table {%
6.875 -8.77537994563491e-16
7.125 -8.77537994563491e-16
};
\addplot [black]
table {%
6.875 29.244003361846
7.125 29.244003361846
};
\addplot [black]
table {%
8 9.26718695357709e-16
8 -1.15719295986394e-16
};
\addplot [black]
table {%
8 28.850718107422
8 29.4011286995754
};
\addplot [black]
table {%
7.875 -1.15719295986394e-16
8.125 -1.15719295986394e-16
};
\addplot [black]
table {%
7.875 29.4011286995754
8.125 29.4011286995754
};
\addplot [black]
table {%
9 -2.1408069757483e-16
9 -3.85730986621315e-16
};
\addplot [black]
table {%
9 0.0128832892384189
9 0.0257665784768379
};
\addplot [black]
table {%
8.875 -3.85730986621315e-16
9.125 -3.85730986621315e-16
};
\addplot [black]
table {%
8.875 0.0257665784768379
9.125 0.0257665784768379
};
\addplot [black]
table {%
1 nan
1 nan
};
\addplot [black]
table {%
1 nan
1 nan
};
\addplot [black]
table {%
0.9625 nan
1.0375 nan
};
\addplot [black]
table {%
0.9625 nan
1.0375 nan
};
\addplot [dashed, semithick,gray, line width=2pt]
table {%
1 27.7211477013
9 27.7211477013
};
\addplot [dashed, semithick, lightgray, line width=2pt]
table {%
1 24.6550554292
9 24.6550554292
};
\path [draw=black, fill=blue!80.3921568627451!black]
(axis cs:0.75,26.4184279207911)
--(axis cs:1.25,26.4184279207911)
--(axis cs:1.25,27.3599037582767)
--(axis cs:0.75,27.3599037582767)
--(axis cs:0.75,26.4184279207911)
--cycle;
\path [draw=black, fill=blue!80.3921568627451!black]
(axis cs:1.75,28.7125497249602)
--(axis cs:2.25,28.7125497249602)
--(axis cs:2.25,29.1265431358468)
--(axis cs:1.75,29.1265431358468)
--(axis cs:1.75,28.7125497249602)
--cycle;
\path [draw=black, fill=blue!80.3921568627451!black]
(axis cs:2.75,28.9082253658792)
--(axis cs:3.25,28.9082253658792)
--(axis cs:3.25,29.1772500617182)
--(axis cs:2.75,29.1772500617182)
--(axis cs:2.75,28.9082253658792)
--cycle;
\path [draw=black, fill=blue!80.3921568627451!black]
(axis cs:3.75,28.8767385120864)
--(axis cs:4.25,28.8767385120864)
--(axis cs:4.25,29.1372917647872)
--(axis cs:3.75,29.1372917647872)
--(axis cs:3.75,28.8767385120864)
--cycle;
\path [draw=black, fill=blue!80.3921568627451!black]
(axis cs:4.75,28.7489274929825)
--(axis cs:5.25,28.7489274929825)
--(axis cs:5.25,29.0577826378834)
--(axis cs:4.75,29.0577826378834)
--(axis cs:4.75,28.7489274929825)
--cycle;
\path [draw=black, fill=blue!80.3921568627451!black]
(axis cs:5.75,28.6456487419929)
--(axis cs:6.25,28.6456487419929)
--(axis cs:6.25,29.2407937687996)
--(axis cs:5.75,29.2407937687996)
--(axis cs:5.75,28.6456487419929)
--cycle;
\path [draw=black, fill=blue!80.3921568627451!black]
(axis cs:6.75,12.1605380830799)
--(axis cs:7.25,12.1605380830799)
--(axis cs:7.25,29.1134769860127)
--(axis cs:6.75,29.1134769860127)
--(axis cs:6.75,12.1605380830799)
--cycle;
\path [draw=black, fill=blue!80.3921568627451!black]
(axis cs:7.75,9.26718695357709e-16)
--(axis cs:8.25,9.26718695357709e-16)
--(axis cs:8.25,28.850718107422)
--(axis cs:7.75,28.850718107422)
--(axis cs:7.75,9.26718695357709e-16)
--cycle;
\path [draw=black, fill=blue!80.3921568627451!black]
(axis cs:8.75,-2.1408069757483e-16)
--(axis cs:9.25,-2.1408069757483e-16)
--(axis cs:9.25,0.0128832892384189)
--(axis cs:8.75,0.0128832892384189)
--(axis cs:8.75,-2.1408069757483e-16)
--cycle;
\path [draw=black, fill=color1]
--cycle;
\addplot [color0]
table {%
0.75 26.5623876968039
1.25 26.5623876968039
};
\addplot [color0]
table {%
1.75 28.8167432092145
2.25 28.8167432092145
};
\addplot [color0]
table {%
2.75 29.0842871695968
3.25 29.0842871695968
};
\addplot [color0]
table {%
3.75 28.9320034598554
4.25 28.9320034598554
};
\addplot [color0]
table {%
4.75 28.9303520294123
5.25 28.9303520294123
};
\addplot [color0]
table {%
5.75 28.9840027270199
6.25 28.9840027270199
};
\addplot [color0]
table {%
6.75 28.3492219821834
7.25 28.3492219821834
};
\addplot [color0]
table {%
7.75 2.72310551268741e-05
8.25 2.72310551268741e-05
};
\addplot [color0]
table {%
8.75 -4.24304085283448e-17
9.25 -4.24304085283448e-17
};
\addplot [color0]
table {%
0.925 nan
1.075 nan
};
\end{axis}

\end{tikzpicture}

%% file: arch_comparison_filter_id0.tex
\begin{tikzpicture}

\definecolor{color0}{rgb}{0.203921568627451,0.541176470588235,0.741176470588235}
\definecolor{color1}{rgb}{0,0.749019607843137,1}

\begin{axis}[
scale only axis=true,
width=0.3\textwidth,
height=0.2\textwidth,
axis line style={white},
legend cell align={left},
legend style={at={(0.03,0.97)}, anchor=north west}, 
tick align=outside,
tick pos=left,
xlabel={Max number filter},
xmajorgrids,
xmin=0.5, xmax=7.5,
xtick style={color=white!33.33333333333333!black},
xtick={1,2,3,4,5,6,7,1},
xticklabels={8,16,32,64,128,256,512,},
ylabel={SINR},
ymajorgrids,
ymin=-1.63573474115987, ymax=34.3504295643572,
ytick style={color=white!33.33333333333333!black}
]
\addplot [black, forget plot]
table {%
1 -4.24304085283448e-17
1 -8.77537994563491e-16
};
\addplot [black, forget plot]
table {%
1 29.1852998498678
1 29.4276590833633
};
\addplot [black, forget plot]
table {%
0.875 -8.77537994563491e-16
1.125 -8.77537994563491e-16
};
\addplot [black, forget plot]
table {%
0.875 29.4276590833633
1.125 29.4276590833633
};
\addplot [black, forget plot]
table {%
2 -1.15719295986394e-16
2 -7.13602325249433e-16
};
\addplot [black, forget plot]
table {%
2 28.9320034598554
2 29.3079324478299
};
\addplot [black, forget plot]
table {%
1.875 -7.13602325249433e-16
2.125 -7.13602325249433e-16
};
\addplot [black, forget plot]
table {%
1.875 29.3079324478299
2.125 29.3079324478299
};
\addplot [black, forget plot]
table {%
3 28.6957275512446
3 28.6957275512446
};
\addplot [black, forget plot]
table {%
3 29.0393777317087
3 29.2756456636568
};
\addplot [black, forget plot]
table {%
2.875 28.6957275512446
3.125 28.6957275512446
};
\addplot [black, forget plot]
table {%
2.875 29.2756456636568
3.125 29.2756456636568
};
\addplot [black, mark=*, mark size=3, mark options={solid,fill opacity=0}, only marks, forget plot]
table {%
3 26.3005577890913
3 0.0257665784768379
};
\addplot [black, forget plot]
table {%
4 28.054088735611
4 26.5623876968039
};
\addplot [black, forget plot]
table {%
4 29.1086725257274
4 29.5150946335932
};
\addplot [black, forget plot]
table {%
3.875 26.5623876968039
4.125 26.5623876968039
};
\addplot [black, forget plot]
table {%
3.875 29.5150946335932
4.125 29.5150946335932
};
\addplot [black, mark=*, mark size=3, mark options={solid,fill opacity=0}, only marks, forget plot]
table {%
4 2.72310551268741e-05
};
\addplot [black, forget plot]
table {%
5 26.4351158276677
5 24.3210761661598
};
\addplot [black, forget plot]
table {%
5 28.9765743172574
5 29.1883591341982
};
\addplot [black, forget plot]
table {%
4.875 24.3210761661598
5.125 24.3210761661598
};
\addplot [black, forget plot]
table {%
4.875 29.1883591341982
5.125 29.1883591341982
};
\addplot [black, mark=*, mark size=3, mark options={solid,fill opacity=0}, only marks, forget plot]
table {%
5 9.27683022824262e-16
};
\addplot [black, forget plot]
table {%
6 28.1459279193399
6 27.5800118017163
};
\addplot [black, forget plot]
table {%
6 29.0155840170641
6 29.3939849094197
};
\addplot [black, forget plot]
table {%
5.875 27.5800118017163
6.125 27.5800118017163
};
\addplot [black, forget plot]
table {%
5.875 29.3939849094197
6.125 29.3939849094197
};
\addplot [black, forget plot]
table {%
7 28.8060690398296
7 28.3159371967028
};
\addplot [black, forget plot]
table {%
7 29.2063322970307
7 29.4011286995754
};
\addplot [black, forget plot]
table {%
6.875 28.3159371967028
7.125 28.3159371967028
};
\addplot [black, forget plot]
table {%
6.875 29.4011286995754
7.125 29.4011286995754
};
\addplot [black, mark=*, mark size=3, mark options={solid,fill opacity=0}, only marks, forget plot]
table {%
7 32.7146948231973
};
\addplot [black, forget plot]
table {%
1 nan
1 nan
};
\addplot [black, forget plot]
table {%
1 nan
1 nan
};
\addplot [black, forget plot]
table {%
0.9625 nan
1.0375 nan
};
\addplot [black, forget plot]
table {%
0.9625 nan
1.0375 nan
};
\addplot [dashed, semithick, gray, line width=2pt]
table {%
1 27.7211477013
7 27.7211477013
};
\addplot [dashed, semithick, lightgray, line width=2pt]
table {%
1 24.6550554292
7 24.6550554292
};
\path [draw=black, fill=blue!80.3921568627451!black]
(axis cs:0.75,-4.24304085283448e-17)
--(axis cs:1.25,-4.24304085283448e-17)
--(axis cs:1.25,29.1852998498678)
--(axis cs:0.75,29.1852998498678)
--(axis cs:0.75,-4.24304085283448e-17)
--cycle;
\path [draw=black, fill=blue!80.3921568627451!black]
(axis cs:1.75,-1.15719295986394e-16)
--(axis cs:2.25,-1.15719295986394e-16)
--(axis cs:2.25,28.9320034598554)
--(axis cs:1.75,28.9320034598554)
--(axis cs:1.75,-1.15719295986394e-16)
--cycle;
\path [draw=black, fill=blue!80.3921568627451!black]
(axis cs:2.75,28.6957275512446)
--(axis cs:3.25,28.6957275512446)
--(axis cs:3.25,29.0393777317087)
--(axis cs:2.75,29.0393777317087)
--(axis cs:2.75,28.6957275512446)
--cycle;
\path [draw=black, fill=blue!80.3921568627451!black]
(axis cs:3.75,28.054088735611)
--(axis cs:4.25,28.054088735611)
--(axis cs:4.25,29.1086725257274)
--(axis cs:3.75,29.1086725257274)
--(axis cs:3.75,28.054088735611)
--cycle;
\path [draw=black, fill=blue!80.3921568627451!black]
(axis cs:4.75,26.4351158276677)
--(axis cs:5.25,26.4351158276677)
--(axis cs:5.25,28.9765743172574)
--(axis cs:4.75,28.9765743172574)
--(axis cs:4.75,26.4351158276677)
--cycle;
\path [draw=black, fill=blue!80.3921568627451!black]
(axis cs:5.75,28.1459279193399)
--(axis cs:6.25,28.1459279193399)
--(axis cs:6.25,29.0155840170641)
--(axis cs:5.75,29.0155840170641)
--(axis cs:5.75,28.1459279193399)
--cycle;
\path [draw=black, fill=blue!80.3921568627451!black]
(axis cs:6.75,28.8060690398296)
--(axis cs:7.25,28.8060690398296)
--(axis cs:7.25,29.2063322970307)
--(axis cs:6.75,29.2063322970307)
--(axis cs:6.75,28.8060690398296)
--cycle;
\path [draw=black, fill=color1]
--cycle;
\addplot [color0, forget plot]
table {%
0.75 25.9940390959043
1.25 25.9940390959043
};
\addplot [color0, forget plot]
table {%
1.75 28.3985515113048
2.25 28.3985515113048
};
\addplot [color0, forget plot]
table {%
2.75 28.9303520294123
3.25 28.9303520294123
};
\addplot [color0, forget plot]
table {%
3.75 28.9385716660725
4.25 28.9385716660725
};
\addplot [color0, forget plot]
table {%
4.75 28.8172094090224
5.25 28.8172094090224
};
\addplot [color0, forget plot]
table {%
5.75 28.4257905511872
6.25 28.4257905511872
};
\addplot [color0, forget plot]
table {%
6.75 28.9685495333987
7.25 28.9685495333987
};
\addplot [color0, forget plot]
table {%
0.925 nan
1.075 nan
};
\end{axis}

\end{tikzpicture}

%% file: arch_comparison_samples_mean_id0.tex
\begin{tikzpicture}

\definecolor{color0}{rgb}{0,0.749019607843137,1}
\definecolor{color1}{rgb}{0,0.75,0}
\definecolor{color3}{rgb}{0.63, 0.79, 0.95}
\definecolor{color2}{rgb}{0,0,0.796875}

\begin{axis}[
scale only axis=true,
width=0.6\textwidth,
height=0.2\textwidth,
axis line style={white},
legend cell align={left},
legend entries={{Sim},{Sim+VReal},{Real},{Transfer},{Interfered},{Clean}},
legend style={at={(0.98,0.62)}, anchor=north east}, 
tick align=outside,
tick pos=left,
xmajorgrids,
xmin=22.5, xmax=627.5,
xtick style={color=white!33.33333333333333!black},
ylabel={Mean SINR},
ymajorgrids,
ymin=7.15480662997358, ymax=33.7441063159219,
ytick style={color=white!33.33333333333333!black}
]

\addlegendimage{solid, no markers, color3, line width=2pt}
\addlegendimage{solid, no markers, color0, line width=2pt}
\addlegendimage{solid, no markers, color2, line width=2pt}
\addlegendimage{solid, no markers, color1, line width=2pt}
\addlegendimage{dashed, no markers, lightgray, line width=2pt}
\addlegendimage{dashed, no markers, gray, line width=2pt}

\addplot [dashed, semithick, gray, line width=2pt]
table {%
	50 27.7211477013
	100 27.7211477013
	150 27.7211477013
	200 27.7211477013
	250 27.7211477013
	300 27.7211477013
	350 27.7211477013
	400 27.7211477013
	450 27.7211477013
	500 27.7211477013
	600 27.7211477013
};

\addplot [dashed, semithick, lightgray, line width=2pt]
table {%
	50 24.6550554292
	100 24.6550554292
	150 24.6550554292
	200 24.6550554292
	250 24.6550554292
	300 24.6550554292
	350 24.6550554292
	400 24.6550554292
	450 24.6550554292
	500 24.6550554292
	600 24.6550554292
};

\addplot [semithick, color0, line width=2pt]
table {%
50 28.2313716014349
100 28.8515016079907
150 29.2032927786103
200 29.114405468307
250 29.354624793851
300 29.1092224097171
350 29.1249430034977
400 29.0245736516827
450 29.5378905179666
500 29.2587905769831
550 29.2979454090671
600 29.3330409582276
};

\addplot [semithick, color3, line width=2pt]
table {%
50 28.3029308920486
100 28.9702565500223
150 29.2117777552212
200 29.3186715821425
250 29.380277214027
300 29.4170176709305
350 29.3392140412506
400 29.3631191147481
450 29.5783322956504
500 29.3399164907064
550 29.4830761081339
600 29.5099176130082
};

\addplot [semithick, color2, line width=2pt]
table {%
50 10.1166394130269
100 14.1672230320882
150 13.9291942172938
200 15.6966338681522
250 18.9816597004227
300 18.2378995777047
350 23.8869932193305
400 26.4809160480598
450 30.1309847353862
500 31.5614874380711
550 31.8044386472017
600 32.1408852720725
};

\addplot [semithick, color1, line width=2pt]
table {%
50 13.7295826613843
100 14.0739334611992
150 14.1299691616005
200 19.0650569457614
250 12.2655813580797
300 19.0973936978574
350 23.2560095947773
400 26.7521664643302
450 28.3085432580813
500 31.6693235213231
550 30.8811694032571
600 32.1126941092845
};

\end{axis}

\end{tikzpicture}

%% file: arch_comparison_samples_var_id0.tex
\begin{tikzpicture}

\definecolor{color0}{rgb}{0,0.749019607843137,1}
\definecolor{color1}{rgb}{0,0.75,0}
\definecolor{color3}{rgb}{0.63, 0.79, 0.95}
\definecolor{color2}{rgb}{0,0,0.796875}

\begin{axis}[
scale only axis=true,
width=0.6\textwidth,
height=0.2\textwidth,
axis line style={white},
tick align=outside,
tick pos=left,
xlabel={Number samples},
xmajorgrids,
xmin=22.5, xmax=627.5,
xtick style={color=white!33.33333333333333!black},
ylabel={Variance of SINR},
ymajorgrids,
ymin=-8.91065807724068, ymax=187.212454889855,
ytick style={color=white!33.33333333333333!black}
]

\addplot [semithick, color0, line width=2pt]
table {%
50 0.00450254028093874
100 0.033214785749818
150 0.00798491638600858
200 0.0228924199842149
250 0.00580097200501891
300 0.0848063770099561
350 0.0186048145910014
400 0.0622548475107068
450 0.00846870864026002
500 0.04284526426485
550 0.0133898378558401
600 0.0526637153991573
};

\addplot [semithick, color3, line width=2pt]
table {%
50 0.0102535900571007
100 0.00397003811941863
150 0.00764438920680381
200 0.0117941179500347
250 0.0101003768979973
300 0.021012282300516
350 0.0211593568524219
400 0.00836265453955553
450 0.0125407176677344
500 0.00402887580913201
550 0.0173664153613724
600 0.0368680304395673
};

\addplot [semithick, color2, line width=2pt]
table {%
50 14.0219765213578
100 67.421242367108
150 95.7497144230179
200 109.656994451846
250 164.506467476792
300 161.920706765671
350 101.713916353621
400 34.0107838269982
450 8.87698143826296
500 3.72975560507608
550 3.26308861055721
600 2.14233949111513
};

\addplot [semithick, color1, line width=2pt]
table {%

50 47.8782011058854
100 69.1177985793885
150 83.2688610512465
200 126.180312222837
250 136.132477528441
300 178.297767936805
350 97.8282795994227
400 56.5239256506149
450 62.2369961705505
500 2.17746660509472
550 2.63376995281693
600 3.09998161039415
};
\end{axis}

\end{tikzpicture}

%% file: cdf_classical.tex
\begin{tikzpicture}


\definecolor{color2}{rgb}{0.95,0.58,0.027} 
\definecolor{color1}{rgb}{0,0,0.796875} 
\definecolor{color3}{rgb}{0.61, 0.023, 0.023} 
\definecolor{color0}{rgb}{0.84,0.31,0.07} 

\begin{axis}[
scale only axis=true,
width=\columnwidth,
height=0.48\columnwidth,
axis line style={white},
legend cell align={left},
legend entries={{Zeroing},{IMAT},{Ramp filtering},{CNN},{Interfered},{Clean}},
legend style={at={(0.97,0.03)}, anchor=south east, draw=white!80.0!black, font=\scriptsize}, 
tick align=outside,
tick pos=left,
xmajorgrids,
xmin=7.26130570160447, xmax=49.725619118139,
ymajorgrids,
ymin=-0.049475, ymax=1.049975,
ytick={-0.2,0,0.2,0.4,0.6,0.8,1,1.2},
yticklabels={−0.2,0.0,0.2,0.4,0.6,0.8,1.0,1.2},
xlabel={$x$ [dB]},
ylabel={Probability P(SINR $\leq$ $x$)}
]

\addlegendimage{solid, no markers, color3, line width=2pt}
\addlegendimage{solid, no markers, color0, line width=2pt}
\addlegendimage{solid, no markers, color2, line width=2pt}
\addlegendimage{solid, no markers, color1, line width=2pt}
\addlegendimage{dashed, no markers, lightgray, line width=2pt}
\addlegendimage{dashed, no markers, gray, line width=2pt}

\addplot [dashed, lightgray, line width=2pt]
table {%
	9.1915017659924 0.004
	9.79017144856273 0.008
	10.6371147894726 0.012
	11.4756651964848 0.016
	12.0721257114237 0.02
	12.2416888644037 0.024
	13.492704192958 0.028
	13.8662522735922 0.032
	14.6815177468483 0.036
	16.2689802776558 0.04
	16.3047215029971 0.044
	16.6152602029074 0.048
	16.9890967936127 0.052
	17.0694335444923 0.056
	17.273640329938 0.06
	17.4650579644238 0.064
	17.8968177298443 0.068
	18.1942576334459 0.072
	18.2231499133276 0.076
	18.2500161494962 0.08
	18.3527144676882 0.084
	18.574647483743 0.088
	18.7071744287735 0.092
	18.7321999033791 0.096
	18.8846328656198 0.1
	18.9120157834671 0.104
	19.0922219516835 0.108
	19.0983629309837 0.112
	19.1735090784049 0.116
	19.3198449528815 0.12
	19.4075583175948 0.124
	19.4757591110915 0.128
	19.5202959442796 0.132
	19.6172543251452 0.136
	19.6535484205271 0.14
	19.7905958557788 0.144
	19.8470208226026 0.148
	19.9369669790003 0.152
	20.0609845070755 0.156
	20.0965423401871 0.16
	20.1188093856576 0.164
	20.1922028359003 0.168
	20.3163534084992 0.172
	20.5670961014692 0.176
	20.6592446058418 0.18
	20.8454878926183 0.184
	20.9700225010706 0.188
	21.2857514634558 0.192
	21.3057867477134 0.196
	21.4089475275589 0.2
	21.4092715917429 0.204
	21.4099655842471 0.208
	21.5037813636218 0.212
	21.6329408011949 0.216
	21.6507486797618 0.22
	21.8128869779696 0.224
	21.852695936545 0.228
	21.8890502672743 0.232
	21.9254556461682 0.236
	21.931820694074 0.24
	21.9597867064655 0.244
	22.0979029885178 0.248
	22.1082003811637 0.252
	22.186768043086 0.256
	22.1959650041975 0.26
	22.2076177265984 0.264
	22.2774904078086 0.268
	22.3518000962121 0.272
	22.3631331579683 0.276
	22.4104682163911 0.28
	22.6762853875414 0.284
	22.7264635987752 0.288
	22.7485027944415 0.292
	22.7875466525594 0.296
	22.7992557618778 0.3
	22.8660197568272 0.304
	22.9232464062935 0.308
	22.9621191951557 0.312
	23.0366763669438 0.316
	23.1623056116141 0.32
	23.1909437768031 0.324
	23.194991914238 0.328
	23.2839357551584 0.332
	23.2927341457422 0.336
	23.3141578938205 0.34
	23.3790908650791 0.344
	23.4800765089947 0.348
	23.5199446018334 0.352
	23.5335780293108 0.356
	23.7808235015998 0.36
	23.7998651363419 0.364
	23.8392583165672 0.368
	23.8537036381817 0.372
	23.8651412185994 0.376
	23.8651622213281 0.38
	23.9189130933943 0.384
	23.9842425987876 0.388
	24.015791996439 0.392
	24.0416758385914 0.396
	24.0540851818691 0.4
	24.1009136928736 0.404
	24.1446376060761 0.408
	24.2811996137326 0.412
	24.2846455216866 0.416
	24.3056908101229 0.42
	24.3902288913553 0.424
	24.4154530350415 0.428
	24.5073824685175 0.432
	24.517250677372 0.436
	24.5471689746871 0.44
	24.5472261479468 0.444
	24.5904639462347 0.448
	24.6287348766981 0.452
	24.6823093830654 0.456
	24.7394400375664 0.46
	24.7569426573019 0.464
	24.8291436980832 0.468
	24.8565241273443 0.472
	24.887617593038 0.476
	24.9216073172844 0.48
	24.9383571592337 0.484
	25.0255598299969 0.488
	25.0356550255006 0.492
	25.0515764924912 0.496
	25.0670540492738 0.5
	25.0765336342831 0.504
	25.1227487421052 0.508
	25.1570871438402 0.512
	25.1720961418878 0.516
	25.2529511104721 0.52
	25.2688531725423 0.524
	25.3199115572553 0.528
	25.3286032001838 0.532
	25.3945220458747 0.536
	25.4054571340539 0.54
	25.5751510190113 0.544
	25.575222804284 0.548
	25.62486294091 0.552
	25.6384160252182 0.556
	25.6433698885475 0.56
	25.6546238119792 0.564
	25.6747170840733 0.568
	25.7314375807318 0.572
	25.7460217070121 0.576
	25.7851911791476 0.58
	25.8031602358131 0.584
	25.814929347157 0.588
	25.8446441644546 0.592
	25.8553658981213 0.596
	25.8919746945763 0.6
	25.895011235522 0.604
	25.9558037911421 0.608
	26.0388683762675 0.612
	26.0486319995635 0.616
	26.072802641113 0.62
	26.1373681905908 0.624
	26.1808930895545 0.628
	26.2367263146563 0.632
	26.2656826312237 0.636
	26.299298803201 0.64
	26.3155428016262 0.644
	26.3553270665993 0.648
	26.4684665177888 0.652
	26.6387998717198 0.656
	26.6525822607347 0.66
	26.7503715637723 0.664
	26.7850193508655 0.668
	26.7902999981889 0.672
	26.8199490532577 0.676
	26.8432247386843 0.68
	26.8476419290916 0.684
	26.9320285903238 0.688
	26.9770437999373 0.692
	27.0595316972864 0.696
	27.0787520373731 0.7
	27.1568276069575 0.704
	27.2282634188312 0.708
	27.2695466451965 0.712
	27.3525543394071 0.716
	27.366765833071 0.72
	27.5039389601215 0.724
	27.5576809280538 0.728
	27.5839861516126 0.732
	27.6037703248346 0.736
	27.64539387021 0.74
	27.6795784041084 0.744
	27.681158407083 0.748
	27.6909156333349 0.752
	27.79512556626 0.756
	28.0328045332502 0.76
	28.1714053989865 0.764
	28.2222476168444 0.768
	28.3448090191672 0.772
	28.5024679677824 0.776
	28.5062493938257 0.78
	28.5208903976813 0.784
	28.5936032196972 0.788
	28.6281046007254 0.792
	28.7049403451176 0.796
	28.7299996725336 0.8
	28.7464290494771 0.804
	28.7759312272362 0.808
	28.8115448506236 0.812
	28.865675310918 0.816
	28.9238494739541 0.82
	28.9744694648377 0.824
	29.0847083729573 0.828
	29.2817835137872 0.832
	29.4434908862282 0.836
	29.4925631440131 0.84
	29.5748393412906 0.844
	29.6704543391086 0.848
	29.736503215836 0.852
	29.8570635446337 0.856
	29.9153048805494 0.86
	29.9703722393692 0.864
	29.9796168711559 0.868
	30.0439570288072 0.872
	30.0498640611631 0.876
	30.1248926067746 0.88
	30.3619018498298 0.884
	30.4109519550643 0.888
	30.4399042025277 0.892
	30.5567403596668 0.896
	30.8604682306588 0.9
	31.0101089783603 0.904
	31.1217715729949 0.908
	31.2572698054339 0.912
	31.6174548260298 0.916
	31.6836814317873 0.92
	31.9066658041041 0.924
	32.0765867750435 0.928
	32.107075767227 0.932
	32.1474090560115 0.936
	32.205227715762 0.94
	32.2216129546241 0.944
	32.3242234693564 0.948
	32.5076748767062 0.952
	32.5825612581485 0.956
	32.8069207101339 0.96
	32.9141195906449 0.964
	32.9416219573547 0.968
	33.0565163266781 0.972
	33.1136782517857 0.976
	33.7303509074689 0.98
	34.4963358288404 0.984
	36.4529376545133 0.988
	36.7895291721331 0.992
	38.7420570710412 0.996
	42.2161621303257 1
};

\addplot [dashed, gray, line width=2pt]
table {%
	20.29999141943 0.004
	20.5602965669563 0.008
	20.6582176702484 0.012
	21.0784352707193 0.016
	21.6035463823331 0.02
	21.705458599076 0.024
	22.1189548357617 0.028
	22.2273486519237 0.032
	22.2700933879976 0.036
	22.3525135716262 0.04
	22.4377054206756 0.044
	22.4474518915456 0.048
	22.5466318551608 0.052
	22.6415811457274 0.056
	22.8454901620065 0.06
	22.9644720931708 0.064
	23.20254702948 0.068
	23.2907607123366 0.072
	23.4056734482345 0.076
	23.5113479236408 0.08
	23.5193245724272 0.084
	23.5201929168315 0.088
	23.5608251091898 0.092
	23.6294419054328 0.096
	23.6297720765247 0.1
	23.6523511139388 0.104
	23.6853044693719 0.108
	23.6929047488436 0.112
	23.777150988438 0.116
	23.9705082338298 0.12
	24.0280946760245 0.124
	24.1247371452783 0.128
	24.2161342387477 0.132
	24.2424523327442 0.136
	24.2860094495022 0.14
	24.302631077986 0.144
	24.304643289934 0.148
	24.4582400757475 0.152
	24.4960427654329 0.156
	24.5386369167744 0.16
	24.5707851265401 0.164
	24.6488387992473 0.168
	24.6765905928828 0.172
	24.6875978372916 0.176
	24.7451951147514 0.18
	24.7968535576867 0.184
	24.7979937312377 0.188
	24.8311016418251 0.192
	24.8746499431735 0.196
	24.9125395219681 0.2
	24.9409004285849 0.204
	24.9899975501105 0.208
	25.0137529661574 0.212
	25.0574642234824 0.216
	25.0928578606995 0.22
	25.1373237477456 0.224
	25.1386318454382 0.228
	25.1410680885531 0.232
	25.2008231882176 0.236
	25.2687054470014 0.24
	25.273734777601 0.244
	25.3155077505712 0.248
	25.3273372621173 0.252
	25.3346713393572 0.256
	25.3380456040253 0.26
	25.4717530991747 0.264
	25.5309927424175 0.268
	25.5844224905868 0.272
	25.5971914685192 0.276
	25.6546238119792 0.28
	25.6557375486855 0.284
	25.6641077579007 0.288
	25.6950444508655 0.292
	25.7310284079105 0.296
	25.8306049831857 0.3
	25.8329378064766 0.304
	25.8486105738996 0.308
	25.8992321429564 0.312
	25.9168428378958 0.316
	25.9586704543636 0.32
	25.9648888016152 0.324
	25.9997559919723 0.328
	26.0506307943532 0.332
	26.1854420957228 0.336
	26.1912754156263 0.34
	26.2001652875467 0.344
	26.2262480286634 0.348
	26.230204963557 0.352
	26.2416870481143 0.356
	26.2678914497678 0.36
	26.3101005936427 0.364
	26.3382321458113 0.368
	26.3531922859155 0.372
	26.360047868545 0.376
	26.3699097178986 0.38
	26.4225023079588 0.384
	26.4513418350798 0.388
	26.4668802483062 0.392
	26.4700616590334 0.396
	26.4877717350296 0.4
	26.4968134978297 0.404
	26.4997592798118 0.408
	26.5050667230261 0.412
	26.5098451426324 0.416
	26.5311187271475 0.42
	26.5442114288503 0.424
	26.5586973619455 0.428
	26.6113554404608 0.432
	26.6227666081173 0.436
	26.6477266561787 0.44
	26.6521453266348 0.444
	26.7116566060823 0.448
	26.7448146256445 0.452
	26.7592291825556 0.456
	26.7657105927403 0.46
	26.8066952044777 0.464
	26.8090137833411 0.468
	26.9299720301405 0.472
	26.9571233899727 0.476
	27.0189813045674 0.48
	27.0325781071202 0.484
	27.045451720673 0.488
	27.0929062339015 0.492
	27.1201479949848 0.496
	27.1655517691678 0.5
	27.1681101139433 0.504
	27.1973965384675 0.508
	27.2170582085015 0.512
	27.2540715884839 0.516
	27.3008082613488 0.52
	27.3545439375894 0.524
	27.3780927078411 0.528
	27.3994915827613 0.532
	27.5093235510845 0.536
	27.5919131108224 0.54
	27.5966557765367 0.544
	27.6217312561227 0.548
	27.6379462095851 0.552
	27.6628007144146 0.556
	27.7004659279893 0.56
	27.7514639278771 0.564
	27.7799316580742 0.568
	27.8135149089876 0.572
	27.9219910694749 0.576
	27.9464052566391 0.58
	27.9522535084344 0.584
	28.0356356959175 0.588
	28.0748485202021 0.592
	28.167022816908 0.596
	28.1980239858833 0.6
	28.237748747809 0.604
	28.2917878228893 0.608
	28.3240200341143 0.612
	28.3790767438627 0.616
	28.3938257190829 0.62
	28.4740975815288 0.624
	28.5883814609576 0.628
	28.5938696407752 0.632
	28.6089861004373 0.636
	28.6226393207064 0.64
	28.6473373784285 0.644
	28.6540836442425 0.648
	28.7585324025775 0.652
	28.7748373723202 0.656
	28.7891279529859 0.66
	28.8192113286785 0.664
	28.826340225126 0.668
	28.8371249901656 0.672
	28.8520539894536 0.676
	28.8952288535242 0.68
	28.9000614788279 0.684
	29.0236916231796 0.688
	29.0338111544337 0.692
	29.0602367729833 0.696
	29.2037625951171 0.7
	29.2217069037595 0.704
	29.2248736345704 0.708
	29.4578357224514 0.712
	29.4597254274576 0.716
	29.5343785327868 0.72
	29.6116210740389 0.724
	29.62015651524 0.728
	29.6357136802318 0.732
	29.6585876180648 0.736
	29.6984383529628 0.74
	29.6996258718812 0.744
	29.8255475342399 0.748
	29.8378740901645 0.752
	29.853474626515 0.756
	29.8570586692369 0.76
	29.86423230696 0.764
	29.8798177306485 0.768
	29.8853232779159 0.772
	29.9594833580073 0.776
	29.9771122739116 0.78
	29.9984276648894 0.784
	30.0391676747203 0.788
	30.0669760575663 0.792
	30.1000890169452 0.796
	30.1030192797206 0.8
	30.1153834967049 0.804
	30.2572108635997 0.808
	30.2824288134574 0.812
	30.3295451718887 0.816
	30.4407940895054 0.82
	30.4542342214432 0.824
	30.487542701571 0.828
	30.6424696049098 0.832
	30.7299718802523 0.836
	30.7725501843059 0.84
	30.8439867585941 0.844
	30.9637404326286 0.848
	31.1012421029234 0.852
	31.2178776337196 0.856
	31.3419535578793 0.86
	31.5172336867308 0.864
	31.6350447309958 0.868
	31.8081297513255 0.872
	31.997679031276 0.876
	32.0064181179155 0.88
	32.1438935009058 0.884
	32.1722305301901 0.888
	32.2072913749424 0.892
	32.3625781459407 0.896
	32.3996674414208 0.9
	32.4208296154433 0.904
	32.5203664014406 0.908
	32.7086588070258 0.912
	32.812765078445 0.916
	32.8338108672106 0.92
	32.9315571349905 0.924
	32.9948006832755 0.928
	33.0565163266781 0.932
	33.5806516914188 0.936
	33.6805920228722 0.94
	33.7514032517994 0.944
	33.9277831765479 0.948
	34.213818614141 0.952
	34.7204676176377 0.956
	34.8950926896255 0.96
	34.9340528660178 0.964
	35.0330785368274 0.968
	35.2398501992581 0.972
	36.2803451479761 0.976
	36.6803847798795 0.98
	37.9202408480884 0.984
	38.3687426824474 0.988
	38.8905556422711 0.992
	39.6802382480043 0.996
	42.4069562563969 1
};

\addplot [solid, color3, line width=2pt]
table[row sep=crcr]{%
	15.4894212604441	0.004\\
	15.8985085545593	0.008\\
	15.9251624212333	0.012\\
	16.1585280396984	0.016\\
	16.6461658542699	0.02\\
	16.9760847954249	0.024\\
	17.0987352576037	0.028\\
	17.3406642638189	0.032\\
	17.4366521088433	0.036\\
	17.5348273740707	0.04\\
	17.6726709340128	0.044\\
	17.681536286151	0.048\\
	18.0604866989555	0.052\\
	18.2288404319357	0.056\\
	18.6861156146231	0.06\\
	18.6992062075489	0.064\\
	18.7636972535864	0.068\\
	18.8006152006606	0.072\\
	18.8322288366535	0.076\\
	18.8796791339693	0.08\\
	18.9137264931991	0.084\\
	18.9185883836907	0.088\\
	19.1832517568386	0.092\\
	19.1969192893396	0.096\\
	19.2133884792869	0.1\\
	19.2744050084262	0.104\\
	19.3748322048255	0.108\\
	19.4432347541894	0.112\\
	19.4941641765579	0.116\\
	19.6058848001245	0.12\\
	19.6222414376073	0.124\\
	19.6820058621826	0.128\\
	19.8751642936077	0.132\\
	19.926805823554	0.136\\
	20.009082706051	0.14\\
	20.0413589456058	0.144\\
	20.0470299923017	0.148\\
	20.0971870891506	0.152\\
	20.1143022509753	0.156\\
	20.1369480690938	0.16\\
	20.2110383389909	0.164\\
	20.3527809701462	0.168\\
	20.3747274481093	0.172\\
	20.3937846281133	0.176\\
	20.4323033548073	0.18\\
	20.4896717944674	0.184\\
	20.5234536281068	0.188\\
	20.5553391601486	0.192\\
	20.6250774618423	0.196\\
	20.7078883130042	0.2\\
	20.758269783575	0.204\\
	20.7634101433124	0.208\\
	20.7659259368011	0.212\\
	20.796125891993	0.216\\
	20.8085757156909	0.22\\
	20.8336061799255	0.224\\
	20.8740306306643	0.228\\
	20.9348968589768	0.232\\
	21.042158045447	0.236\\
	21.0628299548248	0.24\\
	21.0872318959993	0.244\\
	21.0954349713651	0.248\\
	21.109008329391	0.252\\
	21.1122848335379	0.256\\
	21.1158117473573	0.26\\
	21.1441126450264	0.264\\
	21.1830619263327	0.268\\
	21.188448169163	0.272\\
	21.1915035953317	0.276\\
	21.1922193145391	0.28\\
	21.1971842012469	0.284\\
	21.2026377498457	0.288\\
	21.2200678402734	0.292\\
	21.2261658365971	0.296\\
	21.2280944276575	0.3\\
	21.3605902581486	0.304\\
	21.3626818135467	0.308\\
	21.3687476362687	0.312\\
	21.3928428458299	0.316\\
	21.4067616765913	0.32\\
	21.4233114009352	0.324\\
	21.4960268494459	0.328\\
	21.5165109536258	0.332\\
	21.5405547450625	0.336\\
	21.669462929127	0.34\\
	21.700182961226	0.344\\
	21.7273238523744	0.348\\
	21.7560007006633	0.352\\
	21.7658478010594	0.356\\
	21.8388977271611	0.36\\
	21.8930847169452	0.364\\
	21.8953200091036	0.368\\
	21.8993767569358	0.372\\
	21.9142397758017	0.376\\
	21.9286714610794	0.38\\
	21.9417115891137	0.384\\
	21.9562417528903	0.388\\
	21.9863750347588	0.392\\
	22.0420276605047	0.396\\
	22.0504159185343	0.4\\
	22.0935332776517	0.404\\
	22.1683225122042	0.408\\
	22.169864223871	0.412\\
	22.2218573855031	0.416\\
	22.235305982497	0.42\\
	22.2498852105273	0.424\\
	22.2545286994299	0.428\\
	22.3848722170182	0.432\\
	22.4076084411985	0.436\\
	22.5424990345127	0.44\\
	22.5787709640985	0.444\\
	22.665150247458	0.448\\
	22.6660033185633	0.452\\
	22.7050099309613	0.456\\
	22.7129807399651	0.46\\
	22.7132263962691	0.464\\
	22.7305224040419	0.468\\
	22.7719893143266	0.472\\
	22.9089008400463	0.476\\
	22.9099362249655	0.48\\
	22.9300785825062	0.484\\
	22.9754681106266	0.488\\
	23.0003311490076	0.492\\
	23.0452408909	0.496\\
	23.0471952446996	0.5\\
	23.056147080915	0.504\\
	23.0585414106434	0.508\\
	23.0668449000423	0.512\\
	23.1179574683614	0.516\\
	23.1408631870125	0.52\\
	23.1838761402506	0.524\\
	23.2078555053301	0.528\\
	23.3712991483519	0.532\\
	23.4161984138278	0.536\\
	23.4939277178745	0.54\\
	23.592618983415	0.544\\
	23.6003708704751	0.548\\
	23.6228967181901	0.552\\
	23.6387160274539	0.556\\
	23.6466724289635	0.56\\
	23.6638690226863	0.564\\
	23.6975985319254	0.568\\
	23.7154068583077	0.572\\
	23.7499676171814	0.576\\
	23.7825286583721	0.58\\
	24.0629680151838	0.584\\
	24.1396993625684	0.588\\
	24.2615269821712	0.592\\
	24.3271331158626	0.596\\
	24.4307871828894	0.6\\
	24.4703936204323	0.604\\
	24.4746848579869	0.608\\
	24.5706290028916	0.612\\
	24.6552957871478	0.616\\
	24.7490481662809	0.62\\
	24.8255474285127	0.624\\
	24.8259118504067	0.628\\
	24.8503951978041	0.632\\
	24.98152518436	0.636\\
	24.9948095498027	0.64\\
	25.0105671758271	0.644\\
	25.0481331345796	0.648\\
	25.0611626079545	0.652\\
	25.0715004919577	0.656\\
	25.0781131043069	0.66\\
	25.0989275075622	0.664\\
	25.1007098977065	0.668\\
	25.1423076139084	0.672\\
	25.1508519116369	0.676\\
	25.2270668685028	0.68\\
	25.2526879217552	0.684\\
	25.2771627690476	0.688\\
	25.3239828624075	0.692\\
	25.4560683453561	0.696\\
	25.4839704428877	0.7\\
	25.5126063654335	0.704\\
	25.517835356541	0.708\\
	25.5354958435362	0.712\\
	25.5667381692219	0.716\\
	25.5784833199744	0.72\\
	25.6381149731766	0.724\\
	25.8393744203829	0.728\\
	25.8556310277295	0.732\\
	25.8824736803745	0.736\\
	25.9323982303933	0.74\\
	25.9516205758652	0.744\\
	26.1788366255355	0.748\\
	26.2126207767938	0.752\\
	26.2298072686556	0.756\\
	26.3505125869542	0.76\\
	26.3702978356067	0.764\\
	26.4934276126291	0.768\\
	26.5737176005557	0.772\\
	26.661162485623	0.776\\
	26.6628149785347	0.78\\
	26.6685895779892	0.784\\
	26.8733125192443	0.788\\
	26.8891373797233	0.792\\
	26.926716742884	0.796\\
	26.9867490824009	0.8\\
	27.0021462732191	0.804\\
	27.0789179482446	0.808\\
	27.1025876305198	0.812\\
	27.2112921691382	0.816\\
	27.2610035988839	0.82\\
	27.2848983358769	0.824\\
	27.2921940643594	0.828\\
	27.3429915092928	0.832\\
	27.3650435979632	0.836\\
	27.4095090831311	0.84\\
	27.4347929606017	0.844\\
	27.4769012508482	0.848\\
	27.4779132074335	0.852\\
	27.5517819479533	0.856\\
	27.6168255998972	0.86\\
	27.6585695474806	0.864\\
	27.6777982518085	0.868\\
	27.7616116023595	0.872\\
	27.7724483497304	0.876\\
	27.7893251658458	0.88\\
	28.015363254745	0.884\\
	28.1240282642608	0.888\\
	28.210342273345	0.892\\
	28.3882831291724	0.896\\
	28.4594851842212	0.9\\
	28.6882574485722	0.904\\
	28.7921900723358	0.908\\
	28.8048417438069	0.912\\
	28.8354711360929	0.916\\
	28.8494482724765	0.92\\
	28.9322986544809	0.924\\
	29.3131236036499	0.928\\
	29.5680616005372	0.932\\
	29.6694902126826	0.936\\
	29.7241102451123	0.94\\
	29.7940184310615	0.944\\
	29.9764335500741	0.948\\
	30.517187892252	0.952\\
	30.7861993003178	0.956\\
	31.0469551095507	0.96\\
	31.5458955877694	0.964\\
	31.6844668677341	0.968\\
	31.7405119747982	0.972\\
	31.9500910583137	0.976\\
	32.7169795406276	0.98\\
	33.3066440272022	0.984\\
	33.7777810623989	0.988\\
	34.6257256876753	0.992\\
	35.1498479783096	0.996\\
	35.8862335654952	1\\
};

\addplot [solid, color0, line width=2pt]
table[row sep=crcr]{%
	15.4580031771457	0.004\\
	15.8487633529975	0.008\\
	15.9105212477715	0.012\\
	16.8632612057087	0.016\\
	17.4679192823703	0.02\\
	17.5037014306847	0.024\\
	17.5441561823665	0.028\\
	17.823981537394	0.032\\
	17.8370997278953	0.036\\
	18.0072595440017	0.04\\
	18.1028470693471	0.044\\
	18.2048799802348	0.048\\
	18.2291243600134	0.052\\
	18.5265411311864	0.056\\
	18.5924744415963	0.06\\
	18.6728566393799	0.064\\
	18.8765136296223	0.068\\
	18.9340680608258	0.072\\
	18.996146562723	0.076\\
	19.087710935115	0.08\\
	19.1282528241175	0.084\\
	19.294956786939	0.088\\
	19.4205192354839	0.092\\
	19.5134960954182	0.096\\
	19.6372660733545	0.1\\
	19.6829667573476	0.104\\
	19.7206707339705	0.108\\
	19.7593449346677	0.112\\
	19.7826149767863	0.116\\
	19.8223340439571	0.12\\
	19.8560556949765	0.124\\
	19.8726647832678	0.128\\
	19.9190831567114	0.132\\
	20.065223359835	0.136\\
	20.2003324226998	0.14\\
	20.2243673723015	0.144\\
	20.2492542176637	0.148\\
	20.2904413535756	0.152\\
	20.3094725716718	0.156\\
	20.4240999147983	0.16\\
	20.5181315141478	0.164\\
	20.6791895627864	0.168\\
	20.6798419893707	0.172\\
	20.7269011406196	0.176\\
	20.7272168367678	0.18\\
	20.7418886493458	0.184\\
	20.7617206883543	0.188\\
	20.7803289078708	0.192\\
	20.7915258140863	0.196\\
	20.8259380480737	0.2\\
	20.8629447089691	0.204\\
	20.8853921182935	0.208\\
	20.9252602066685	0.212\\
	20.9445772466292	0.216\\
	20.9607737329364	0.22\\
	20.9753059428862	0.224\\
	20.9914785751856	0.228\\
	21.0005708888622	0.232\\
	21.0454778671427	0.236\\
	21.0790580948856	0.24\\
	21.1079785020935	0.244\\
	21.1266712834907	0.248\\
	21.1379321244273	0.252\\
	21.1408794761416	0.256\\
	21.1992485860536	0.26\\
	21.2122154923207	0.264\\
	21.230659766156	0.268\\
	21.2653708466342	0.272\\
	21.2783438550039	0.276\\
	21.2827317686632	0.28\\
	21.297676992832	0.284\\
	21.3266471325705	0.288\\
	21.3346213935701	0.292\\
	21.3576277722887	0.296\\
	21.4048574684903	0.3\\
	21.4535892824026	0.304\\
	21.4632541023129	0.308\\
	21.4686939690556	0.312\\
	21.4970865009117	0.316\\
	21.5017174858647	0.32\\
	21.6401351250201	0.324\\
	21.6763068063932	0.328\\
	21.6765363683329	0.332\\
	21.6927548949091	0.336\\
	21.6959860911313	0.34\\
	21.7088417657896	0.344\\
	21.7149596342265	0.348\\
	21.7515908419193	0.352\\
	21.7935993230616	0.356\\
	21.8633980632347	0.36\\
	21.9100753098367	0.364\\
	21.9140057286199	0.368\\
	22.0648578350665	0.372\\
	22.1521452832026	0.376\\
	22.1574760742621	0.38\\
	22.1773398711482	0.384\\
	22.2148195779877	0.388\\
	22.2332518016886	0.392\\
	22.2633282139068	0.396\\
	22.2726807765811	0.4\\
	22.3378277739849	0.404\\
	22.3536941187053	0.408\\
	22.3688149345246	0.412\\
	22.5099014294251	0.416\\
	22.5749850544001	0.42\\
	22.6343999983968	0.424\\
	22.6410432660109	0.428\\
	22.7031713078893	0.432\\
	22.8008790389801	0.436\\
	22.819160715681	0.44\\
	22.9020373322166	0.444\\
	22.9543734084674	0.448\\
	22.958079170449	0.452\\
	22.9731658913485	0.456\\
	23.0129854048755	0.46\\
	23.0318113151074	0.464\\
	23.0896066922967	0.468\\
	23.0897743147037	0.472\\
	23.11354456731	0.476\\
	23.1607129287241	0.48\\
	23.1644648606409	0.484\\
	23.1669517148273	0.488\\
	23.1797591314979	0.492\\
	23.2090657165595	0.496\\
	23.2713110502475	0.5\\
	23.3213741295858	0.504\\
	23.3807845816072	0.508\\
	23.3878278893026	0.512\\
	23.3926723657335	0.516\\
	23.4117068264017	0.52\\
	23.4429693708565	0.524\\
	23.4845166153181	0.528\\
	23.5510718909392	0.532\\
	23.6073950534039	0.536\\
	23.6336764429466	0.54\\
	23.6387160274539	0.544\\
	23.6614538419299	0.548\\
	23.6927845283067	0.552\\
	23.7062852055943	0.556\\
	23.7998375727674	0.56\\
	23.847514892241	0.564\\
	23.9125362825823	0.568\\
	23.9663218384152	0.572\\
	24.0785098791243	0.576\\
	24.2541857081433	0.58\\
	24.355667755163	0.584\\
	24.3640397324252	0.588\\
	24.4455916159134	0.592\\
	24.4538479066353	0.596\\
	24.6219932466708	0.6\\
	24.6935777668575	0.604\\
	24.7472636284213	0.608\\
	24.7747272709051	0.612\\
	24.7920407471298	0.616\\
	24.8365319537244	0.62\\
	24.8458473943492	0.624\\
	24.8682894383282	0.628\\
	24.9400237423221	0.632\\
	24.9405211694246	0.636\\
	24.9560713006179	0.64\\
	24.9565678323893	0.644\\
	24.9755681979516	0.648\\
	25.0032055241041	0.652\\
	25.1419450838643	0.656\\
	25.1513474594994	0.66\\
	25.1523976939855	0.664\\
	25.1978115045004	0.668\\
	25.2667777292354	0.672\\
	25.2988742376596	0.676\\
	25.29950015775	0.68\\
	25.3450642772766	0.684\\
	25.4353818972811	0.688\\
	25.5073863485117	0.692\\
	25.5330442611189	0.696\\
	25.5978344224033	0.7\\
	25.660554454704	0.704\\
	25.729127605923	0.708\\
	25.7592342425285	0.712\\
	25.7814564474723	0.716\\
	25.8118396663899	0.72\\
	25.830396703745	0.724\\
	25.8368069596264	0.728\\
	25.9576220447166	0.732\\
	25.9824058205664	0.736\\
	26.010068314578	0.74\\
	26.0890338791789	0.744\\
	26.2206251715811	0.748\\
	26.2749578853804	0.752\\
	26.3665377765214	0.756\\
	26.3952761953304	0.76\\
	26.4104566226967	0.764\\
	26.4497710280126	0.768\\
	26.7147216994573	0.772\\
	26.7627896245583	0.776\\
	26.8663910479436	0.78\\
	26.959518548879	0.784\\
	26.9681090491346	0.788\\
	27.0041854172892	0.792\\
	27.0697684866319	0.796\\
	27.3115874253297	0.8\\
	27.3116739226677	0.804\\
	27.3178116840141	0.808\\
	27.3368813774117	0.812\\
	27.4825455750175	0.816\\
	27.4874925821593	0.82\\
	27.5050852529194	0.824\\
	27.5193693528454	0.828\\
	27.5498108825942	0.832\\
	27.5513124830641	0.836\\
	27.5530814964813	0.84\\
	27.5612938552479	0.844\\
	27.6168661786699	0.848\\
	27.6585695474806	0.852\\
	27.6849043445139	0.856\\
	27.8382661072638	0.86\\
	27.8886307221464	0.864\\
	27.9012205187953	0.868\\
	27.9156987849397	0.872\\
	27.9618408492995	0.876\\
	28.0114201248319	0.88\\
	28.4662569137557	0.884\\
	28.6408837449178	0.888\\
	28.7297360427192	0.892\\
	28.8331932026266	0.896\\
	28.8360091444881	0.9\\
	28.8599927004073	0.904\\
	28.9355301965331	0.908\\
	28.9465766897908	0.912\\
	29.0585224076167	0.916\\
	29.186979551132	0.92\\
	29.2350900025707	0.924\\
	29.2815482334866	0.928\\
	29.3784690404863	0.932\\
	29.504969165659	0.936\\
	29.7388031873294	0.94\\
	29.7675269176793	0.944\\
	29.9571278256663	0.948\\
	30.4822177879731	0.952\\
	30.7712453800724	0.956\\
	31.0584438531609	0.96\\
	31.4846111048073	0.964\\
	31.5316344472482	0.968\\
	31.7968890067768	0.972\\
	32.0189121154134	0.976\\
	33.3294923709168	0.98\\
	33.6047655953723	0.984\\
	34.2065425197047	0.988\\
	34.6000206191968	0.992\\
	35.4496049669643	0.996\\
	36.0596644749673	1\\
};

\addplot [solid, color2, line width=2pt]
table[row sep=crcr]{%
	18.8509727808016	0.004\\
	18.9364272871207	0.008\\
	19.1071265222772	0.012\\
	19.417467936569	0.016\\
	19.7059197477409	0.02\\
	19.9476219691646	0.024\\
	20.1063552558167	0.028\\
	20.4676642567021	0.032\\
	20.4904750572415	0.036\\
	20.5821609919651	0.04\\
	20.6173129180235	0.044\\
	20.7744155981306	0.048\\
	20.8031010245784	0.052\\
	20.8430377297544	0.056\\
	20.9343001762707	0.06\\
	20.9518117654782	0.064\\
	20.9934625543216	0.068\\
	21.0161493446746	0.072\\
	21.0262646486634	0.076\\
	21.0357897517872	0.08\\
	21.1442664320711	0.084\\
	21.1565091917715	0.088\\
	21.1928438738713	0.092\\
	21.2902707275835	0.096\\
	21.4029424603658	0.1\\
	21.409069159731	0.104\\
	21.4412481699699	0.108\\
	21.5652435343671	0.112\\
	21.6232459683342	0.116\\
	21.8546467764953	0.12\\
	21.9260638694335	0.124\\
	21.9332485125857	0.128\\
	21.9641166211572	0.132\\
	22.0195310329582	0.136\\
	22.0578083623018	0.14\\
	22.0703636970136	0.144\\
	22.1091630515772	0.148\\
	22.1157469766268	0.152\\
	22.1249785847473	0.156\\
	22.2748940328189	0.16\\
	22.2989960513108	0.164\\
	22.4067946727975	0.168\\
	22.4472767745632	0.172\\
	22.4552924971742	0.176\\
	22.5298971932725	0.18\\
	22.5950917137837	0.184\\
	22.5987872328878	0.188\\
	22.6035257294798	0.192\\
	22.6564224591452	0.196\\
	22.7932820928177	0.2\\
	22.8001782041165	0.204\\
	22.813269950557	0.208\\
	22.8406762116305	0.212\\
	22.8475550169663	0.216\\
	22.8635279391396	0.22\\
	22.9107966920715	0.224\\
	22.9163580689561	0.228\\
	22.9196532685829	0.232\\
	22.9197156374867	0.236\\
	22.9977376634438	0.24\\
	23.0018128227272	0.244\\
	23.0375296091983	0.248\\
	23.112674060285	0.252\\
	23.1139841745383	0.256\\
	23.1194796887331	0.26\\
	23.1277048304181	0.264\\
	23.150265774842	0.268\\
	23.1830530773521	0.272\\
	23.2139117524834	0.276\\
	23.2620681325539	0.28\\
	23.2786488525639	0.284\\
	23.3290934505085	0.288\\
	23.3490568047085	0.292\\
	23.4408761627679	0.296\\
	23.4442359867711	0.3\\
	23.4444865032803	0.304\\
	23.4789384295628	0.308\\
	23.5019904481533	0.312\\
	23.514476914398	0.316\\
	23.5249926269548	0.32\\
	23.531406707135	0.324\\
	23.5810129048097	0.328\\
	23.6039416241964	0.332\\
	23.6094730996481	0.336\\
	23.6194877009872	0.34\\
	23.6457613414486	0.344\\
	23.6481891641708	0.348\\
	23.7191406287052	0.352\\
	23.7334018329695	0.356\\
	23.7515875571722	0.36\\
	23.7576417585663	0.364\\
	23.7756346553001	0.368\\
	23.778901229973	0.372\\
	23.8051215988651	0.376\\
	23.8414852035922	0.38\\
	23.9051485482451	0.384\\
	23.9321572117721	0.388\\
	23.9338162397638	0.392\\
	23.9346556162739	0.396\\
	23.9354333827834	0.4\\
	23.9495249223344	0.404\\
	23.9960078937814	0.408\\
	24.0062778378486	0.412\\
	24.0154260724298	0.416\\
	24.0603366700173	0.42\\
	24.1342459064752	0.424\\
	24.1395736764795	0.428\\
	24.195146470728	0.432\\
	24.320022322645	0.436\\
	24.34508819418	0.44\\
	24.3457409878465	0.444\\
	24.3556527090484	0.448\\
	24.3636084180681	0.452\\
	24.3673245021665	0.456\\
	24.3811073119046	0.46\\
	24.4360928527361	0.464\\
	24.4541652034948	0.468\\
	24.4757294608587	0.472\\
	24.5001282483545	0.476\\
	24.5549076630846	0.48\\
	24.5552552411747	0.484\\
	24.5803547594774	0.488\\
	24.6173543666861	0.492\\
	24.620599447019	0.496\\
	24.6329502142695	0.5\\
	24.6602421947877	0.504\\
	24.6704747086081	0.508\\
	24.6782677276343	0.512\\
	24.7273085220379	0.516\\
	24.7731857281684	0.52\\
	24.8005421456127	0.524\\
	24.8160876741971	0.528\\
	24.8635621769714	0.532\\
	24.8695867943894	0.536\\
	24.9791844708774	0.54\\
	24.9908468711857	0.544\\
	25.0421077004205	0.548\\
	25.0481008112467	0.552\\
	25.0891425036448	0.556\\
	25.1165254662399	0.56\\
	25.1619678015987	0.564\\
	25.2254751182798	0.568\\
	25.2595343192809	0.572\\
	25.2764997164525	0.576\\
	25.3577382990115	0.58\\
	25.3945632211644	0.584\\
	25.4865266878532	0.588\\
	25.4869390936901	0.592\\
	25.5116469388883	0.596\\
	25.5277378871693	0.6\\
	25.5299085497408	0.604\\
	25.5388540288466	0.608\\
	25.5869965845731	0.612\\
	25.6748856297613	0.616\\
	25.7499421879142	0.62\\
	25.7561412334324	0.624\\
	25.7902928931214	0.628\\
	25.8158026986094	0.632\\
	25.8413742683669	0.636\\
	25.9073797052994	0.64\\
	25.9180435971569	0.644\\
	25.9503043243843	0.648\\
	25.9527766892076	0.652\\
	25.9881970338037	0.656\\
	26.0167745992628	0.66\\
	26.0203304532393	0.664\\
	26.0254653535186	0.668\\
	26.0262496996045	0.672\\
	26.0332490741377	0.676\\
	26.0546330014941	0.68\\
	26.0576677464685	0.684\\
	26.0758055320732	0.688\\
	26.1026933233974	0.692\\
	26.2624237046928	0.696\\
	26.3917772401295	0.7\\
	26.4251476849785	0.704\\
	26.4803644814845	0.708\\
	26.5114645994701	0.712\\
	26.5363098060467	0.716\\
	26.6075038192429	0.72\\
	26.6474869426828	0.724\\
	26.6517632697063	0.728\\
	26.7327812999895	0.732\\
	26.7688569931427	0.736\\
	26.8092602837411	0.74\\
	26.8785358129513	0.744\\
	26.9115240049811	0.748\\
	26.9282812099326	0.752\\
	27.0002612535396	0.756\\
	27.0856642816712	0.76\\
	27.1329993519666	0.764\\
	27.1524415046956	0.768\\
	27.1970521876873	0.772\\
	27.1977515437625	0.776\\
	27.2309746551726	0.78\\
	27.2336118689221	0.784\\
	27.3152930510335	0.788\\
	27.3891050088308	0.792\\
	27.438866667247	0.796\\
	27.4483258786765	0.8\\
	27.6165812988454	0.804\\
	27.7429078967702	0.808\\
	27.7478174514681	0.812\\
	27.8044878445339	0.816\\
	27.8877900854889	0.82\\
	27.9442711839229	0.824\\
	27.9523902018166	0.828\\
	28.0090271284801	0.832\\
	28.1124748660763	0.836\\
	28.224841721632	0.84\\
	28.3536089901033	0.844\\
	28.3604906962542	0.848\\
	28.4326651936939	0.852\\
	28.4687070508526	0.856\\
	28.5629787637153	0.86\\
	28.709066692873	0.864\\
	28.7930933415897	0.868\\
	28.9143984348375	0.872\\
	28.9435112212611	0.876\\
	29.0462033267894	0.88\\
	29.0478059183598	0.884\\
	29.0659665462716	0.888\\
	29.1175631515663	0.892\\
	29.1333702232062	0.896\\
	29.1651472711275	0.9\\
	29.190988981875	0.904\\
	29.318203355683	0.908\\
	29.3394993860647	0.912\\
	29.3746922776841	0.916\\
	29.4076993434385	0.92\\
	29.5784352879443	0.924\\
	29.6258477280681	0.928\\
	30.2789179132359	0.932\\
	30.3340382635664	0.936\\
	30.378588114674	0.94\\
	30.4233404484341	0.944\\
	31.0190079343997	0.948\\
	31.06852552818	0.952\\
	31.3283379313825	0.956\\
	31.4352574479325	0.96\\
	31.5701488317372	0.964\\
	31.7706954291455	0.968\\
	31.7737028408883	0.972\\
	32.2320360787185	0.976\\
	32.9971028640472	0.98\\
	34.107018463407	0.984\\
	34.3101093209897	0.988\\
	34.4006601765412	0.992\\
	34.6910603333616	0.996\\
	36.3427357037728	1\\
};

\addplot [semithick, color1, line width=2pt]
table {%
	13.2959950939258 0.004
	17.4430567610667 0.008
	17.9286258873728 0.012
	19.7374257494386 0.016
	20.6078445348755 0.02
	21.3584325869206 0.024
	21.6756810530454 0.028
	22.5191206898154 0.032
	22.5835448294262 0.036
	23.0176923003861 0.04
	23.4809953211591 0.044
	23.4831743171673 0.048
	23.8986437162087 0.052
	23.90948107893 0.056
	24.575304177715 0.06
	24.7401043309738 0.064
	24.7459998792396 0.068
	24.8463104666023 0.072
	25.1406484255899 0.076
	25.3726849706377 0.08
	25.3900659787932 0.084
	25.5338262815437 0.088
	25.6621608013218 0.092
	25.6913676292993 0.096
	25.9253105524045 0.1
	26.0899699712359 0.104
	26.0940508288885 0.108
	26.2723589315612 0.112
	26.6236292500272 0.116
	26.8483554222681 0.12
	27.0483111373368 0.124
	27.1288229688024 0.128
	27.1567590020746 0.132
	27.1635354578493 0.136
	27.239946091291 0.14
	27.2831654562936 0.144
	27.433861009613 0.148
	27.6140162203493 0.152
	27.7427314312649 0.156
	27.7785950990236 0.16
	28.0566190036252 0.164
	28.1049940518919 0.168
	28.1084507154108 0.172
	28.1159232736067 0.176
	28.1487806302882 0.18
	28.239225120426 0.184
	28.3258682319238 0.188
	28.3816624305823 0.192
	28.4304185570749 0.196
	28.4451370974354 0.2
	28.4580813113499 0.204
	28.4663177815648 0.208
	28.4667961491056 0.212
	28.533997431179 0.216
	28.5440691789518 0.22
	28.5621841006061 0.224
	28.5712328988185 0.228
	28.6045465708235 0.232
	28.6702463626642 0.236
	28.8224972178503 0.24
	28.8944884892353 0.244
	28.9326182076954 0.248
	28.9410287373626 0.252
	29.0045721661103 0.256
	29.0183225945456 0.26
	29.1061885217727 0.264
	29.1231536717204 0.268
	29.1358717935043 0.272
	29.1605262575464 0.276
	29.2493462088207 0.28
	29.2506634562584 0.284
	29.290920562228 0.288
	29.292962711544 0.292
	29.3072296802655 0.296
	29.3584266804949 0.3
	29.4369234885747 0.304
	29.4851685694985 0.308
	29.4984553629321 0.312
	29.528928146677 0.316
	29.5960881578882 0.32
	29.6632200155614 0.324
	29.6789468573119 0.328
	29.681330292769 0.332
	29.6960255296868 0.336
	29.6992174237568 0.34
	29.7821721377463 0.344
	29.7960409266038 0.348
	29.8707147714228 0.352
	29.8952486085359 0.356
	29.9292658239488 0.36
	29.9313038231607 0.364
	30.0118152168508 0.368
	30.0165924849145 0.372
	30.0332469011883 0.376
	30.0630462373077 0.38
	30.0808460172196 0.384
	30.0945810262235 0.388
	30.1005837387132 0.392
	30.1013334953174 0.396
	30.1887498003915 0.4
	30.3215002355749 0.404
	30.3275053883556 0.408
	30.3553001880291 0.412
	30.3921007449235 0.416
	30.4129412697364 0.42
	30.4464901374129 0.424
	30.4754845414781 0.428
	30.4954830448777 0.432
	30.657664123466 0.436
	30.6587472052402 0.44
	30.6808563919005 0.444
	30.6889902339515 0.448
	30.7164265484449 0.452
	30.7441269867015 0.456
	30.7884450505294 0.46
	30.8188912614411 0.464
	30.8269599228265 0.468
	30.8331378249035 0.472
	30.8415761478984 0.476
	30.85123445542 0.48
	31.0688804642344 0.484
	31.160284817936 0.488
	31.1648687416584 0.492
	31.1869137258984 0.496
	31.2367472796697 0.5
	31.4231300746936 0.504
	31.4527640455709 0.508
	31.5610340655891 0.512
	31.5715143067893 0.516
	31.6046016100922 0.52
	31.6098124270583 0.524
	31.6396519893404 0.528
	31.6512769850186 0.532
	31.7145079261064 0.536
	31.749448128252 0.54
	31.7769354465144 0.544
	31.796397869644 0.548
	31.8140181222875 0.552
	31.8341738236064 0.556
	31.8571410918475 0.56
	31.8747375882295 0.564
	31.9341082154719 0.568
	31.9410341148731 0.572
	31.9951076883678 0.576
	32.0711485829286 0.58
	32.1097161133032 0.584
	32.1410347986843 0.588
	32.1830296953915 0.592
	32.1960988947348 0.596
	32.2858816822739 0.6
	32.301385738328 0.604
	32.3726620786899 0.608
	32.3908201393856 0.612
	32.4061666077207 0.616
	32.4071076290644 0.62
	32.4745919027872 0.624
	32.650104420298 0.628
	32.6552511847863 0.632
	32.6996356946536 0.636
	32.7311939723703 0.64
	32.8118843734409 0.644
	32.841873332585 0.648
	32.903745454468 0.652
	32.9130396636374 0.656
	32.9248007721866 0.66
	33.0322058609212 0.664
	33.0917142528942 0.668
	33.1182658288483 0.672
	33.1317157843768 0.676
	33.1770086259466 0.68
	33.187894936837 0.684
	33.322073280002 0.688
	33.3306181376374 0.692
	33.4862550671018 0.696
	33.5335978432986 0.7
	33.5590307427961 0.704
	33.6948884903465 0.708
	33.7834814215165 0.712
	33.9352324315242 0.716
	34.0774583494091 0.72
	34.094463396475 0.724
	34.106758747835 0.728
	34.1098502556195 0.732
	34.1224080688803 0.736
	34.2309025409473 0.74
	34.4167624822879 0.744
	34.4907729586703 0.748
	34.500890896136 0.752
	34.5576513811794 0.756
	34.5869757803177 0.76
	34.5894395696162 0.764
	34.6183546790931 0.768
	34.6502670691913 0.772
	34.8212572749501 0.776
	34.9480111210417 0.78
	35.1779759264853 0.784
	35.2134917931743 0.788
	35.2301674129501 0.792
	35.2454816162707 0.796
	35.2814996354009 0.8
	35.3024174411894 0.804
	35.3422096654029 0.808
	35.4108219137519 0.812
	35.6819531789517 0.816
	35.8669402398884 0.82
	35.9098074056945 0.824
	35.9323183343159 0.828
	36.0480693698952 0.832
	36.1001581322788 0.836
	36.185291002612 0.84
	36.205465776701 0.844
	36.290846173243 0.848
	36.3188295566101 0.852
	36.3217453768974 0.856
	36.5106405815924 0.86
	36.6031461021451 0.864
	36.8265309167917 0.868
	36.9180348965671 0.872
	37.4360507968076 0.876
	37.6143963792673 0.88
	37.6343417035527 0.884
	37.6374234781507 0.888
	37.9418630242643 0.892
	38.1864720904273 0.896
	38.3525796939956 0.9
	38.3808781176901 0.904
	38.3879788870687 0.908
	38.4842709374222 0.912
	38.5001379849074 0.916
	38.7180120686007 0.92
	38.8175618435372 0.924
	39.3328991748714 0.928
	39.4560825129465 0.932
	39.746192198714 0.936
	39.9119181433082 0.94
	39.9924652810724 0.944
	40.2162566515288 0.948
	40.3531056132529 0.952
	40.378954076353 0.956
	40.3844611681355 0.96
	40.9550746261017 0.964
	40.9626762559526 0.968
	41.3615332814179 0.972
	42.085641907433 0.976
	42.103462463927 0.98
	42.8405027773103 0.984
	43.1593477351897 0.988
	43.2495860483743 0.992
	44.2009173290663 0.996
	47.7954230537511 1
};

\end{axis}

\end{tikzpicture}

%% file: eval_evaluation_test_doppler-range_matrix_interfered_p19_id0.tex
\begin{tikzpicture}

\begin{axis}[
scale only axis=true,
width=2.5cm,
height=2.5cm,
axis background/.style={fill=white!89.80392156862746!black},
axis line style={white},
point meta max=0,
point meta min=-30,
tick align=outside,
tick pos=left,
x grid style={white},
xlabel={Velocity [m/s]},
xmajorgrids,
xmin=-15, xmax=14.6703296703297,
xtick style={color=white!33.33333333333333!black},
y grid style={white},
ylabel={Distance [m]},
y label style={at={(axis description cs:0.2,.5)},anchor=south},
ymajorgrids,
ymin=5, ymax=60,
ytick style={color=white!33.33333333333333!black}
]
\addplot graphics [includegraphics cmd=\pgfimage,xmin=-15, xmax=14.6703296703297, ymin=5, ymax=60] {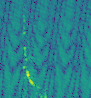};
\end{axis}

\end{tikzpicture}

%% file: eval_evaluation_test_doppler-range_matrix_targets_p19_id0.tex
\begin{tikzpicture}

\begin{axis}[
scale only axis=true,
width=2.5cm,
height=2.5cm,
axis background/.style={fill=white!89.80392156862746!black},
axis line style={white},
point meta max=0,
point meta min=-30,
tick align=outside,
tick pos=left,
x grid style={white},
xlabel={Velocity [m/s]},
xmajorgrids,
xmin=-15, xmax=14.6703296703297,
xtick style={color=white!33.33333333333333!black},
y grid style={white},
yticklabels={,,},
ymajorgrids,
ymin=5, ymax=60,
ytick style={color=white!33.33333333333333!black}
]
\addplot graphics [includegraphics cmd=\pgfimage,xmin=-15, xmax=14.6703296703297, ymin=5, ymax=60] {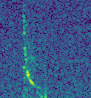};
\end{axis}

\end{tikzpicture}

%% file: eval_evaluation_test_doppler-range_matrix_predictions_p19_id0.tex
\begin{tikzpicture}

\begin{axis}[
scale only axis=true,
width=2.5cm,
height=2.5cm,
axis background/.style={fill=white!89.80392156862746!black},
axis line style={white},
colorbar,
colorbar/width=1.5mm,
colorbar style={ylabel={}},
colormap/viridis,
point meta max=0,
point meta min=-30,
tick align=outside,
tick pos=left,
x grid style={white},
xlabel={Velocity [m/s]},
xmajorgrids,
xmin=-15, xmax=14.6703296703297,
xtick style={color=white!33.33333333333333!black},
y grid style={white},
yticklabels={,,},
ymajorgrids,
ymin=5, ymax=60,
ytick style={color=white!33.33333333333333!black}
]
\addplot graphics [includegraphics cmd=\pgfimage,xmin=-15, xmax=14.6703296703297, ymin=5, ymax=60] {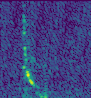};
\end{axis}

\end{tikzpicture}